\documentclass{jfm}
\usepackage{graphicx,epstopdf,epsfig,amsmath,mathtools,subfigure}

\newcommand{\overbar}[1]{\mkern 1.5mu\overline{\mkern-1.5mu#1\mkern-1.5mu}\mkern 1.5mu}
\def\p{\partial}

\shorttitle{Edge state modulation by mean viscosity gradients}
\shortauthor{E. Rinaldi, P. Schlatter and S. Bagheri}

\title{Edge state modulation by mean viscosity gradients}

\author{Enrico Rinaldi\aff{1,\corresp{\email{erinaldi@mech.kth.se}}},
  Philipp Schlatter\aff{1,2}
 \and Shervin Bagheri\aff{1}}

\affiliation{\aff{1}Linn\'e FLOW Centre, KTH Mechanics, Royal Institute of Technology, SE-100 44 Stockholm,
Sweden
\aff{2}Swedish e-Science Research Centre (SeRC), SE-100 44 Stockholm, Sweden}

\begin{document}

\maketitle

\begin{abstract}
Motivated by the relevance of edge state solutions as mediators of transition, we use direct numerical simulations to study the effect of spatially non-uniform viscosity on their energy and stability in minimal channel flows.
What we seek is a theoretical support rooted in a fully non-linear framework that explains the modified threshold for transition to turbulence in flows with temperature-dependent viscosity.
Consistently over a range of subcritical Reynolds numbers, we find that decreasing viscosity away from the walls weakens the streamwise streaks and the vortical structures responsible for their regeneration. 
The entire self-sustained cycle of the edge state is maintained on a lower kinetic energy level with a smaller driving force, compared to a flow with constant viscosity.
Increasing viscosity away from the walls has the opposite effect.
In both cases, the effect is proportional to the strength of the viscosity gradient.
The results presented highlight a local shift in the state space of the position of the edge state relative to the laminar attractor with the consequent modulation of its basin of attraction in the proximity of the edge state and of the surrounding manifold.
The implication is that the threshold for transition is reduced for perturbations evolving in the neighbourhood of the edge state in case viscosity decreases away from the walls, and vice versa.
\end{abstract}

\begin{keywords}
\end{keywords}

\section{Introduction}
\label{sec:intro}

It is well known that transition from a laminar to a turbulent regime in wall-bounded low speed flows is delayed  if the viscosity of the fluid decreases near the solid surfaces, while early transition results from the opposite variation of viscosity.
Experimental evidence was provided since the late 40's by using heated plates in boundary layer flows of air, for which the viscosity increases with temperature. In such flow configuration transition to turbulence occurs earlier, if compared to the unheated case~\citep{1947_LiepmannAndFila}.
On the contrary, an increase of the critical Reynolds number for transition up to an order of magnitude was reported in flows of water, for which the viscosity decreases with temperature, over heated surfaces of various geometries~\citep{1977_StrazisarEtAl,1978_StrazisarAndReshotko,1981_BarkerAndGile,1984_LauchleAndGurney}.
The physical interpretation often invoked to explain this effect is that lower viscosity close to the wall results in a fuller mean velocity profile, which is less susceptible to showing an inflectional point.

Linear stability analysis of the incompressible Navier--Stokes equations with a non-uniform temperature field has been so far the main tool adopted to shed light on how property variations influence transition to turbulence.
Several authors have focused on canonical flows, such as boundary layers or channel flows, and demonstrated the strong stabilising effect of lower viscosity close to the wall on the least damped eigenvalue, if compared to cases with constant properties at the same Reynolds number~\citep{1993_SchaferAndHerwig,1996_WallAndWilson,1997_WallAndWilson}.
\citet{2001_GovindarajanEtAl} showed that the mechanism responsible for the stabilisation of the laminar profile is a reduced intake of energy from the mean flow by the perturbation velocity field, which reduces the production term in the perturbation kinetic energy (PKE) balance.
This effect is maximised when the viscosity variation is localised at the critical layer.

The mentioned studies are based on eigenvalue analyses, thus they provide information on the flow behaviour for long evolution times.
However, transition to turbulence can occur in a subcritical scenario at significantly lower Reynolds numbers than the critical value predicted by the linear theory.
Transient amplification of PKE in linearly stable flows due to the non-normality of the linearized operator offers a possible mechanism to trigger secondary instabilities and, eventually, transition to turbulence~\citep{2001_SchmidAndHenningson}.
\citet{2005_ChikkadiEtAl,2007_SameenAndGovindarajan} have demonstrated that a non-uniform mean viscosity distribution in the direction normal to the walls only marginally affects the maximum transient growth in a plane channel.
This conclusion was reversed by \citet{2007_NouarEtAl}, who accounted for the interaction between viscosity and velocity fluctuations in studying the stability of shear-thinning fluids.
They showed that a viscosity contrast is, indeed, a viable solution to delay transition as the transient energy growth of small perturbations is strongly reduced as the consequence of the interaction between the fluctuating fields.
Further discussion on instabilities of viscosity stratified flows is reported in the review by~\citet{2014_GovindarajanAndSahu}.

Despite the relevance of the large potential for transient growth of perturbations in wall bounded shear flows, transition is a predominantly non-linear phenomenon, whose manifestation depends on the existence of attracting solutions of the Navier--Stokes equations other than the laminar state~\citep{1995_Waleffe}.
In the last decade, new momentum to understanding subcritical transition in linearly stable shear flows has been given by approaching the problem from a fully non-linear perspective rooted in dynamical systems theory.
In a state space representation the laminar flow regime is an attractor while turbulence is generally understood as a saddle.
Experimental and numerical investigations of transition in pipe flows, which are linearly stable to any small perturbations, showed that turbulence is only a transient phenomenon if the Reynolds number is low. 
Localised patches of turbulence, also called puffs, eventually decay towards a laminar state at a rate that decreases super-exponentially with the Reynolds number~\citep{2006_HofEtAl, 2007_EckhardtEtAl, 2008_HofEtAl_PRL}.
Similar observations were documented in Couette flow, where the lifetime of turbulent spots increases with the Reynolds number, yet remaining finite~\citep{2010_SchneiderEtAl_PRE}.
At sufficiently high Reynolds numbers and in large domains puffs start to proliferate at a rate which outpace the decay rate of the single puff and quickly fill the domain bringing the flow to a sustained turbulent state~\citep{2011_AvilaEtAl}.
This scenario changes in small domains, such as the minimal flow unit object of the present study, where turbulent spots do not have enough space to split and expand.
There, turbulence is a saddle even at high Reynolds numbers, however the probability of relaminarisation rapidly drops becoming negligible as this parameter increases.
Regardless of the formal definition of turbulence, a stable manifold exists that separates flow trajectories that relaminarise from the ones that become turbulent, which is commonly referred to as the \textit{edge of chaos}~\citep{2006_SkufcaEtAl,2007_SchneiderEtAl}.
There is now ample evidence that attracting solutions embedded in the manifold, so-called \textit{edge states}, play a crucial role as mediators and harbingers of transition.
These solutions represent minimal self-sustained perturbations that never decay nor become turbulent.

\begin{figure}
\centering
\subfigure[]{\includegraphics[width=0.45\textwidth]{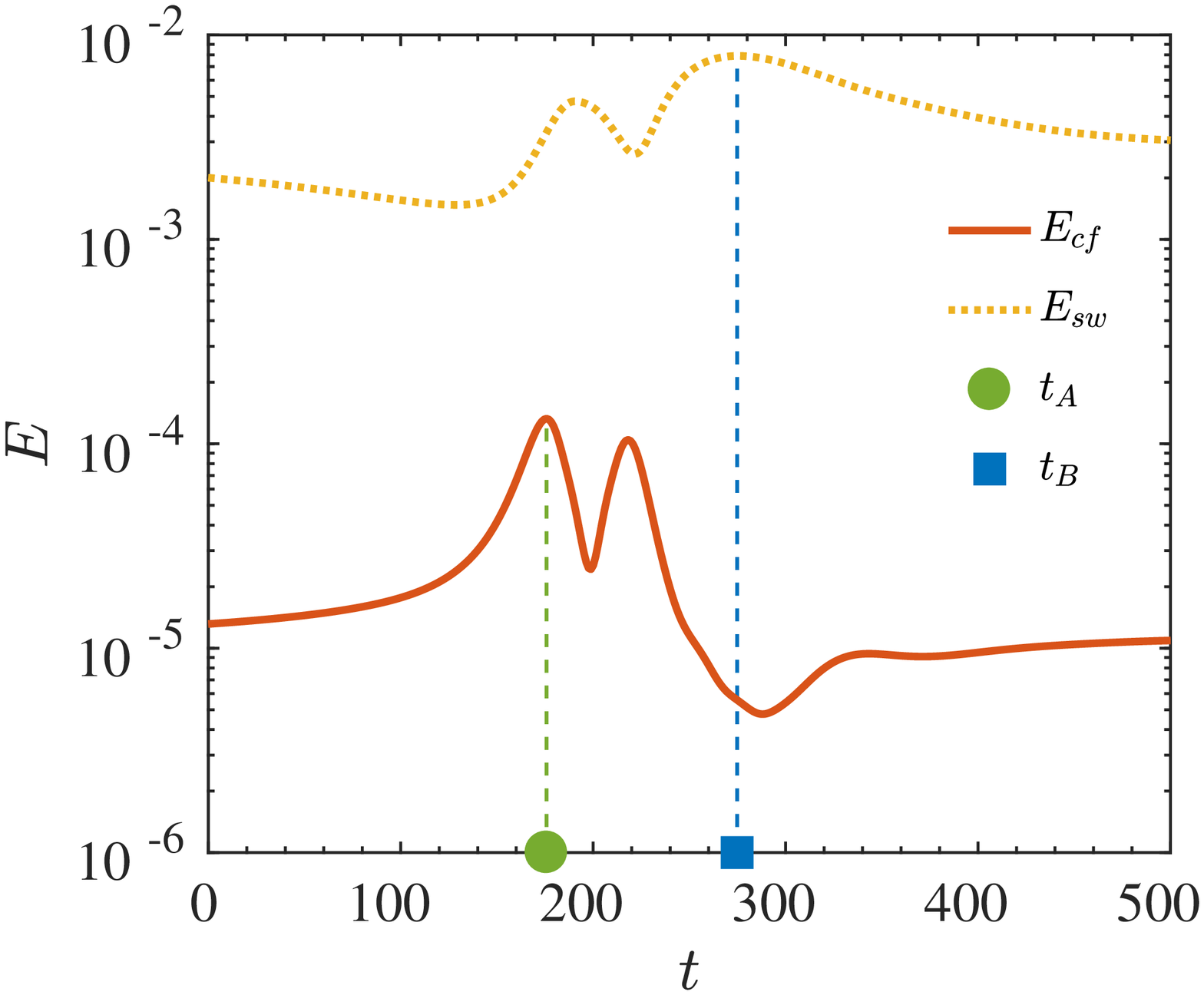}\label{fig:tEn}}
\subfigure[]{\includegraphics[width=0.45\textwidth]{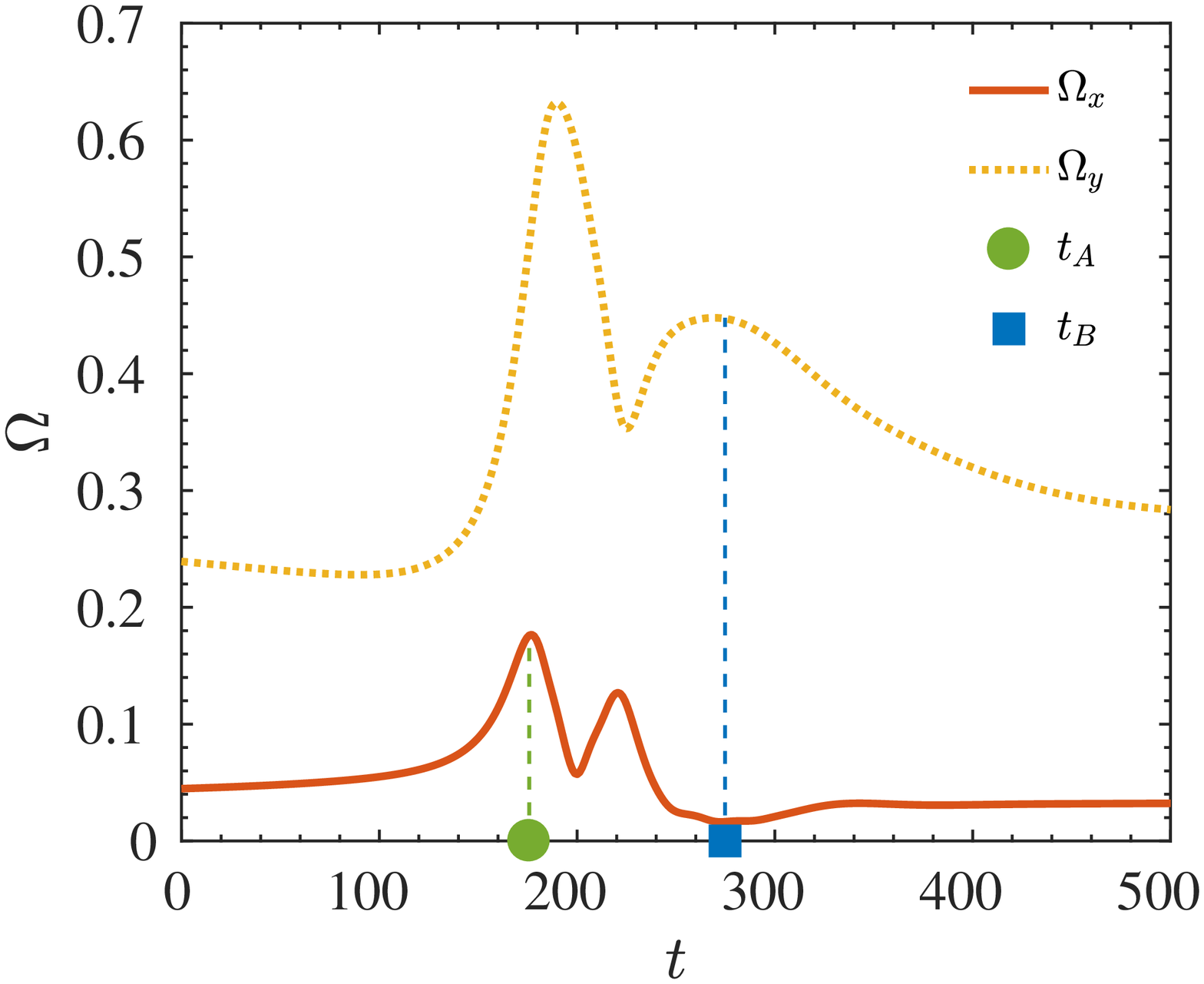}\label{fig:tOm}}
\caption{Time evolution at $Re=2608$ of (a) the streamwise and cross-flow perturbation kinetic energy, $E_{sw}$ and $E_{cf}$, and of (b) the streamwise and wall-normal vorticity, $\Omega_x$ and $\Omega_y$. $t_A$ and $t_B$ indicate the times at which the highest peaks in $E_{cf}$ and $E_{sw}$ occur, respectively, and are the same in figures (a) and (b). At this Reynolds number, the edge state is a relative periodic orbit with period $T=1660$. The definition of perturbation kinetic energy and vorticity is given in \S\ref{sec:numerics}.}
\label{fig:timeplot}
\end{figure}

\begin{figure}
\centering
\subfigure[]{\includegraphics[width=0.32\textwidth]{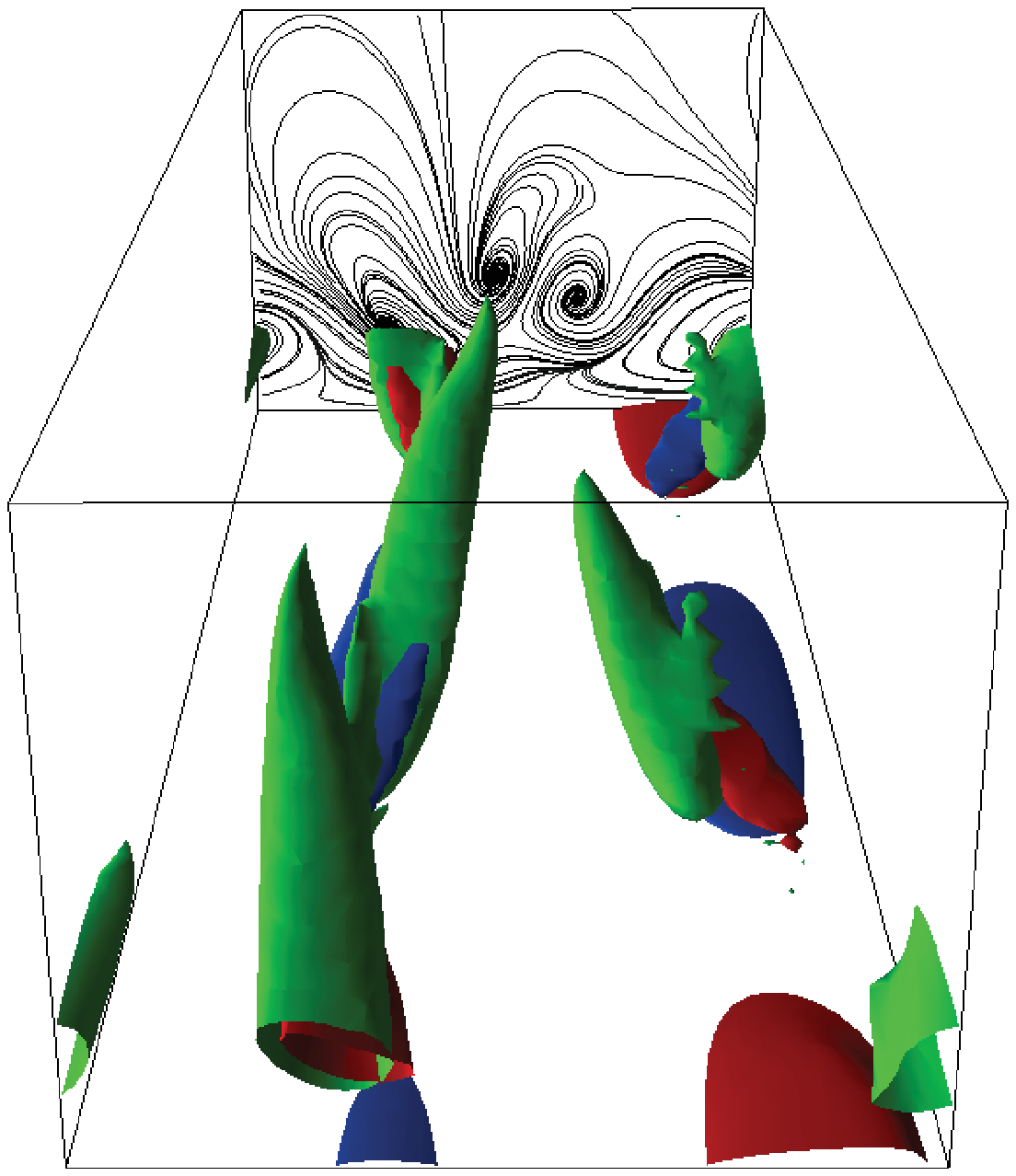}\label{fig:snap_burst}} \quad
\subfigure[]{\includegraphics[width=0.32\textwidth]{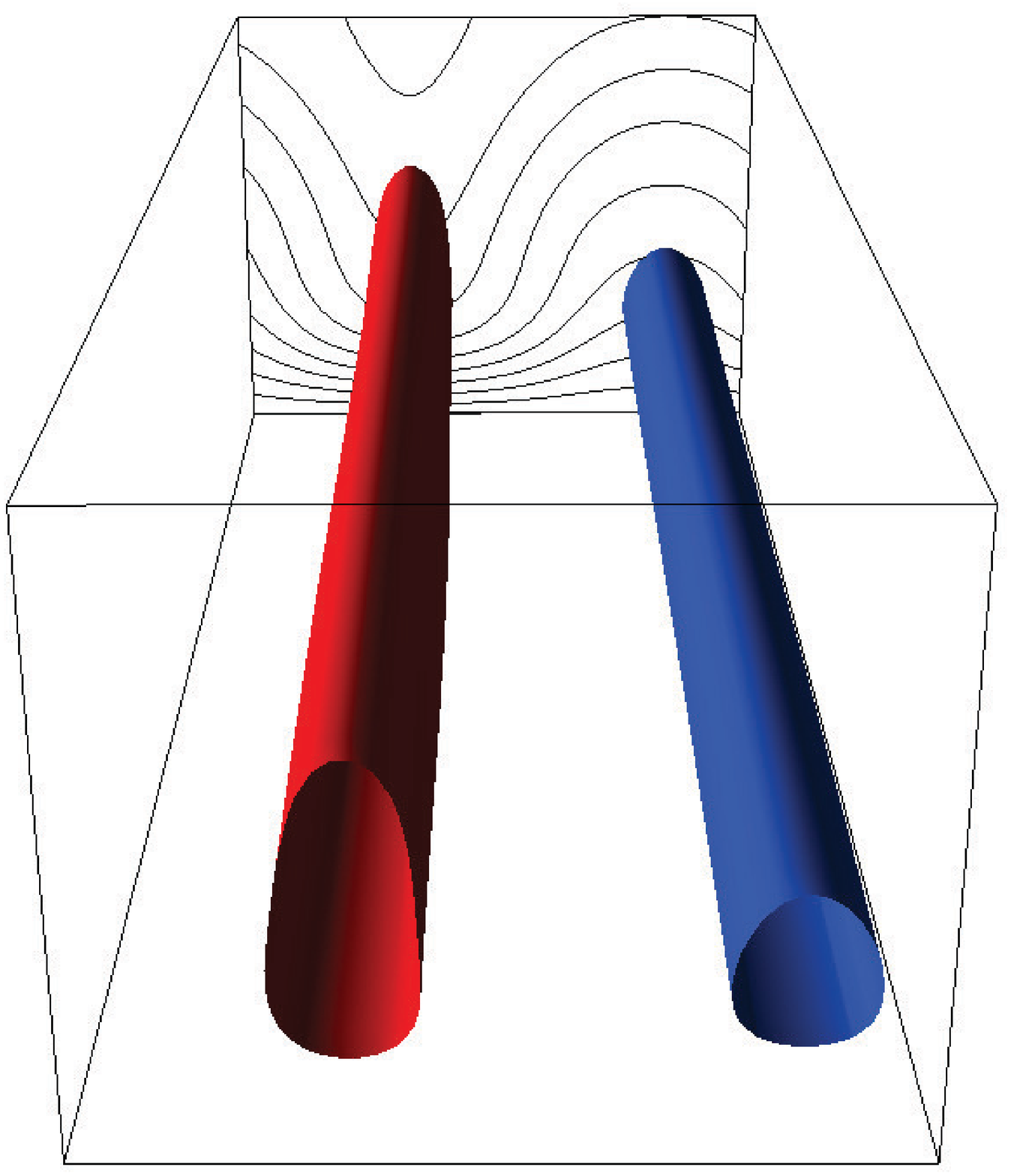}\label{fig:snap_streaks}}
\caption{Snapshots of the flow field at $Re=2608$ at times (a) $t_A$ and (b) $t_B$ as defined in figure~\ref{fig:timeplot}. The edge state is asymmetric and localised in the vertical direction, close to the lower wall, hence only the lower half of the channel is shown. The solid wall is at the bottom and the streamwise flow direction enters the paper in both images. Contours and lines indicate (a) red/blue $\omega_x=\pm 1$, green $\lambda_2=-0.1$ and black lines the streamtracers of the in-plane velocity components, $v$ and $w$; (b) red/blue $u-u_{mean}=\pm 0.2$ and black lines the streamwise velocity profile iso-contours spaced by $\Delta u = 0.11$.}
\label{fig:snapshots}
\end{figure}

\begin{figure}
\centering
\includegraphics[width=0.8\textwidth]{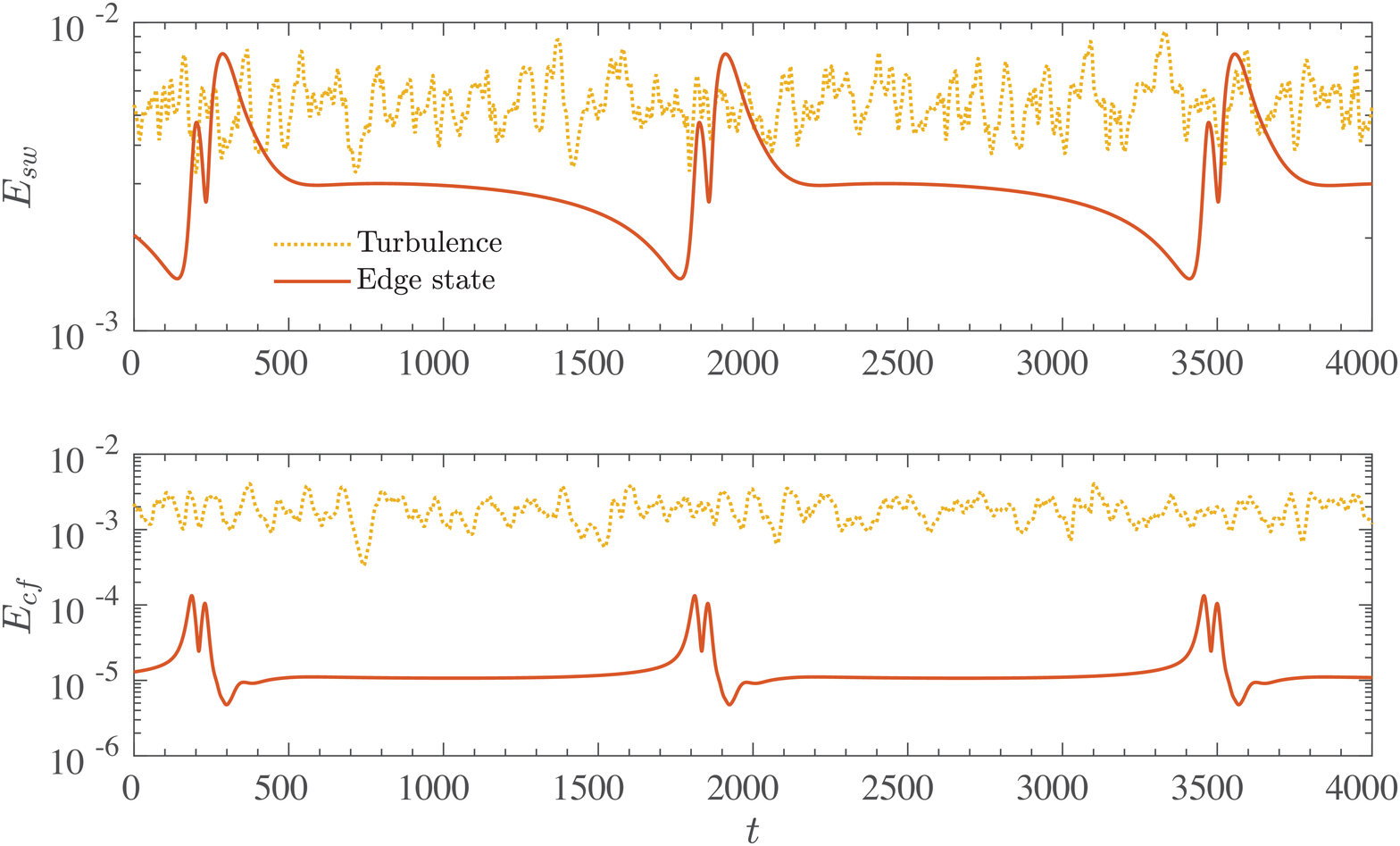}
\caption{Time evolution of the streamwise (top) and cross-flow (bottom) perturbation kinetic energy of the edge state regeneration cycle and of a turbulent trajectory at $Re=2608$. The definition of $E_{sw}$ and $E_{cf}$ is given in \S\ref{sec:numerics}. The energy of the turbulent flow is calculated using the root-mean-square components of the velocity.}
\label{fig:tes}
\end{figure}

Extensive documentation of edge stats is available for several canonical flows, see for example \citet{2008_DuguetEtAl,2008_DuguetEtAl_JFM} for pipe; \citet{2011_CherubiniEtAl,2012_DuguetEtAl,2013_KreilosEtAl,2013_KhapkoEtAl,2016_KhapkoEtAl_JFMR} for boundary layers; \citet{2008_SchneiderEtAl,2009_DuguetEtAl,2013_DuguetEtAl} for Couette; and \citet{2003_TohAndItano,2014_ZammertAndEckhardt,2014_ZammertAndEckhardt_FDR} for channel flows.
Edge states typically appear in the form of travelling waves, relative periodic orbits or chaotic objects, and exhibit low dimensional dynamics of the fluctuating velocity field.
A common feature to all the known solutions is the strong three-dimensional spatial localisation in sufficiently wide and long domains.
On the other hand, in narrow and short domains, such as the minimal flow unit, the localisation is only in the wall-normal direction, see for example the asymmetric flow structures documented in~\citet{2012_XiAndGraham,2014_ZammertAndEckhardt_FDR}.
In such domains, the fluctuating velocity field is reminiscent of turbulence in minimal flow units, where the turbulent fluctuations can be transiently confined to one half-channel only, leaving the other relatively undisturbed~\citep{1991_JimenezAndMoin}.
Despite the lack of localisation in the horizontal directions of edge states in small domains, the active core is qualitatively the same as the one observed in large flow units, see for example~\citep{2013_KhapkoEtAl,2016_KhapkoEtAl_JFMR}.
It consists of flow structures that evolve in a self-sustained cycle closely resembling the one advocated in~\citet{1995_HamiltonEtAl,1997_Waleffe} and the cycle of wall turbulence~\citep{1999_JimenezAndPinelli}.
Three main steps are identified, namely i) formation of streamwise low and high speed streaks; ii) instability of the two dimensional velocity profile; iii) streaks breakup and formation of quasi-streamwise vortices responsible for generating new streaks.
This is exemplified in figures~\ref{fig:timeplot} and~\ref{fig:snapshots} for a nearly minimal channel flow unit at the subcritical Reynolds number $Re=2608$.
In this configuration the edge state is a relative periodic orbit with period $T=1660$.
Projections on the time--energy and time--vorticity planes and snapshots of the perturbation velocity with respect to the mean profile highlight the mentioned steps of the self-sustained process.
In particular, the time at which the streaks reach their maximum amplitude is indicated by $t_B$ and the time at which the intensity of the quasi-streamwise vortices is highest is indicated by $t_A$.
The time evolution of the streamwise and cross-flow perturbation kinetic energy of the edge state regeneration cycle at $Re=2608$ is compared to a turbulent orbit at the same Reynolds number in figure~\ref{fig:tes}.
While the intensity of the streamwise component is comparable between the two regimes, the cross-flow energy of the edge state is on average 2 orders of magnitude smaller.

Transitional flows have been repeatedly shown to approach edge states before crossing the manifold and eventually becoming turbulent~\citep{2009_DuguetEtAl,2009_SchneiderAndEckhardt,2010_SchneiderEtAl,2013_AvilaEtAl_PRL,2016_KhapkoEtAl_JFMR}, and on the path to relaminarisation from a turbulent state~\citep{2012_deLozarEtAl,2015_ParkAndGraham}. 
However, the latter scenario does not constitute the only possible path for relaminarisation~\citep{2014_ChantryAndSchneider}.
Edge states can be regarded as critical nuclei for transition~\citep{2010_SchneiderEtAl} and, being low-branch states, are naturally associated with a threshold in perturbation amplitude~\citep{2008_SchneiderEtAl}.
This conceptual framework was recently used to model the spot nucleation in transitional boundary layers~\citep{2016_KreilosEtAl_PRF}.
Note that, despite their low energy, edge states do not constitute the minimal possible perturbation that triggers transition, also referred to as \textit{minimal seed}, which is identified in the state space by the shortest distance between the laminar attractor and the manifold~\citep{2010_PringleAndKerswell}.
Nonetheless, edge states provide relevant indications on the transition process for perturbations that evolve in their proximity in the state space.
These are not limited to small finite amplitude perturbations carefully generated within a boundary layer, but additionally include a more general scenario of boundary layer receptivity to random noise and spot nucleation~\citep{2016_KhapkoEtAl_JFMR}.

In this paper, we present theoretical arguments stemming from a fully non-linear framework that explain the known modified threshold for transition to turbulence for fluid flows with temperature-dependent viscosity.
As opposed to most previous literature concerned with linearised flow models, we achieve our goal by investigating the modulating effect of an imposed temperature-dependent viscosity profile on edge states, specifically on the perturbation kinetic energy level of their flow structures and on the recurrence of the self-sustained cycle.
Our study follows a similar spirit as~\citet{2010_RolandEtAl}, who showed that in shear-thinning fluids the critical Reynolds number for the appearance of non-linear travelling wave solutions in a pipe is substantially increased, therefore indicating a stabilisation of the flow.
The relevance of our results is in a transition scenario where edge states act as mediators, hence excludes strong perturbations to the flow which possibly bypass their role.
There, a more relevant question is how variable viscosity affects the position and characteristics of the turbulent saddle, which goes beyond the scope of this paper.
The flow configuration considered is the plane channel with a frozen wall-normal symmetric temperature distribution in the absence of gravity, which allows us to isolate the effect of viscosity on the flow.
The validity of the frozen temperature profile for the particular flow case under study is substantiated with \textit{a priori} and \textit{a posteriori} arguments.

\begin{figure}
\centering
\includegraphics[width=0.59\textwidth]{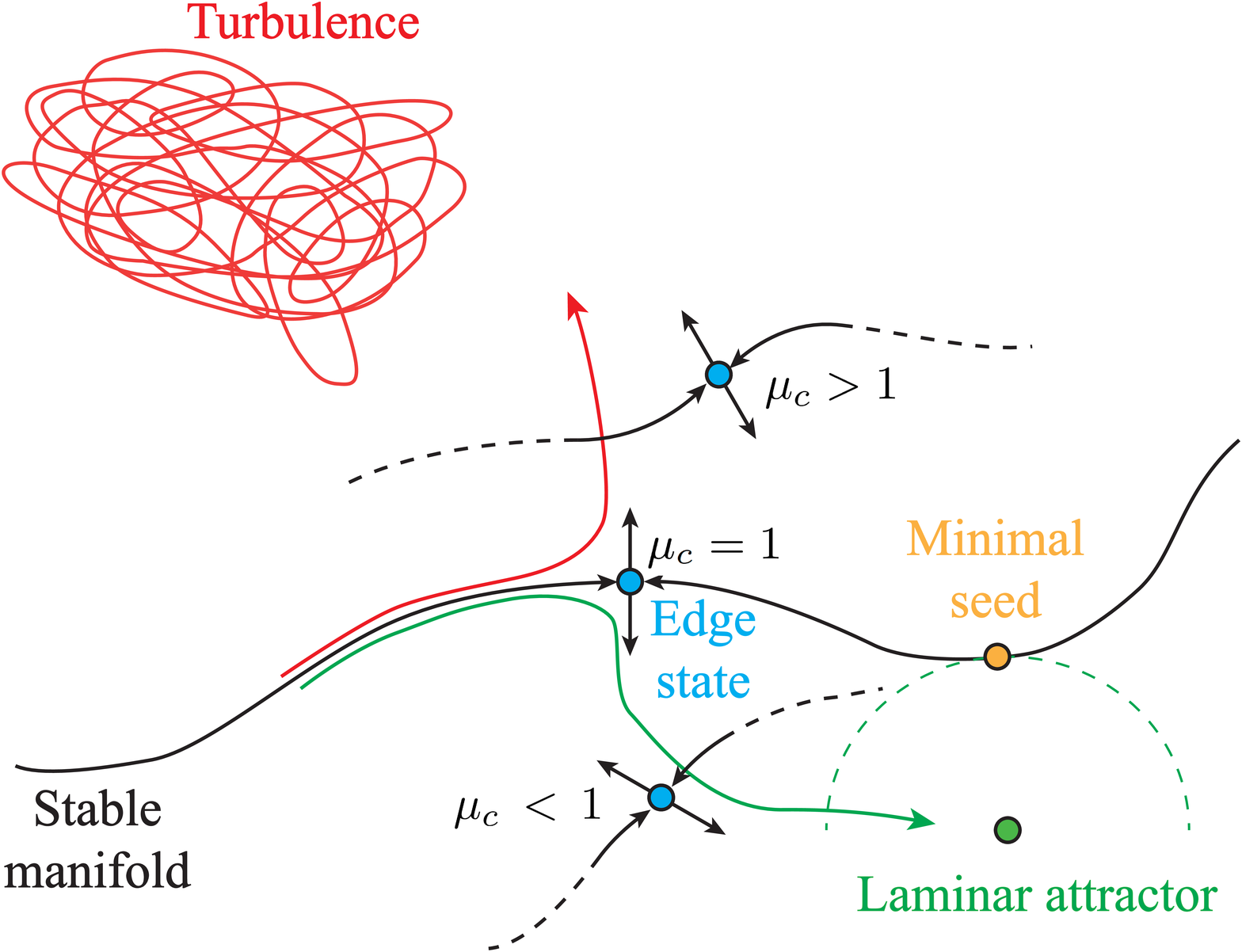}
\caption{Sketch of the state space for a channel flow with unitary non-dimensional reference viscosity at the walls and viscosity at the centerline indicated by $\mu_c$. The variation of viscosity is monotonic towards the centerline according to the constitutive relation given in \S\ref{sec:numerics}. The red and green arrows denote two flow trajectories that become turbulent or relaminarise, respectively. The present results indicate that the position of the edge state shifts away from the laminar attractor, if $\mu_c>1$, or towards it, if $\mu_c>1$. As a consequence, the stable manifold in the vicinity of the edge state is modulated accordingly. This study does not provide a characterisation of the effect of viscosity on the manifold farther away from the edge state (dashed black line). The turbulent saddle is also affected by viscosity.}
\label{fig:sketch0}
\end{figure}

We anticipate the results discussed later by presenting a sketch of the modified state space in figure~\ref{fig:sketch0}.
Smaller viscosity near the walls and larger viscosity at the center of the channel, $\mu_c>1$ in the figure, result in a shift of the position of the edge state away from the laminar attractor and, as a consequence, in a modulation of the the stable manifold in its vicinity.
There, perturbations need to reach larger amplitudes to overcome the local threshold for transition.
The opposite results if viscosity is larger at the walls and smaller at the centerline, $\mu_c<1$.
The described effect is consistent with the behaviour observed in experiments with wall heating or cooling, although it should be noted that the transition scenario taking place in experiments might differ from the one discussed here depending on the specific perturbations applied.

The structure of the paper is as follows. 
\S~\ref{sec:numerics} describes the setup of the problem and the numerical method used in this study.
The main features of edge state solutions in minimal unit channels with constant viscosity are introduced in \S~\ref{sec:constant_visc}.
The effect of viscosity is discussed in \S~\ref{sec:variable_visc}.
Conclusions are presented in \S~\ref{sec:discussion}.

\section{Flow configuration and numerical setup}
\label{sec:numerics}

\subsection{Navier--Stokes equations for a flow with non-uniform mean viscosity}

The configuration studied in this paper is the incompressible flow of a variable viscosity fluid with constant mass flux in a plane channel, for which the dynamics is governed by the Navier--Stokes equations.
In non-dimensional form they read
\begin{eqnarray}\label{eq:NS}
\frac{\partial u_i}{\partial x_i} & = & 0,\\
\frac{\p u_i}{\p t} + u_j\frac{\p u_i}{\p x_j} & = & - \frac{\p p}{\p x_i} + \frac{1}{\overbar{Re}} \frac{\p}{\p x_j} \left( 2 \mu S_{ij} \right).
\end{eqnarray}
In the above expressions $x_i$ indicates the spatial co-ordinates ($x$ streamwise; $y$ wall-normal; $z$ spanwise), $u_i$ is the i-th component of the velocity ($u$ streamwise; $v$ wall-normal; $w$ spanwise), $p$ is the pressure and $S_{ij} = \frac{1}{2}\left( \p u_i / \p x_j + \p u_j / \p x_i \right)$ is the strain rate tensor.
The non-dimensional dynamic viscosity of the fluid is $\mu = \mu_w + \mu_d$, with $\mu_d = \mu_d(y)$ indicating the local deviation from value at the wall, $\mu_w$.
The Reynolds number is defined as $\overbar{Re} = \rho^*_w h \overbar{U}_c / \mu^*_w$, with $\overbar{U}_c$ denoting the centerline velocity, $h$ the half-channel height, and $\rho^*_w$ and $\mu^*_w$ the dimensional density and viscosity reference values, taken as the ones attained at the wall.
Due to the choice of the reference viscosity scale, $\mu_w=1$.
The constitutive relation for the temperature-dependent viscosity mimics the one of a liquid, namely
\begin{equation}\label{eq:visc}
\mu(y) = \frac{1}{\Theta(y)}.
\end{equation}
We impose a symmetric temperature profile that qualitatively resembles the one resulting from a volumetric or wall heating/cooling of a laminar flow with constant thermal conductivity
\begin{equation}\label{eq:temp}
\Theta(y) = 1 + (\Theta_c - 1)(1 - y^2),
\end{equation}
with $\Theta_c$ the centerline value and $y=[-1,1]$. 
Temperature is kept frozen in all simulations, this corresponds to assuming that the interaction between the fluctuating velocity field and the temperature fluctuations is negligible.
This assumption is discussed in more details in \S~\ref{sec:froz_temp}.

\begin{figure}
\centering
\includegraphics[width=0.99\textwidth]{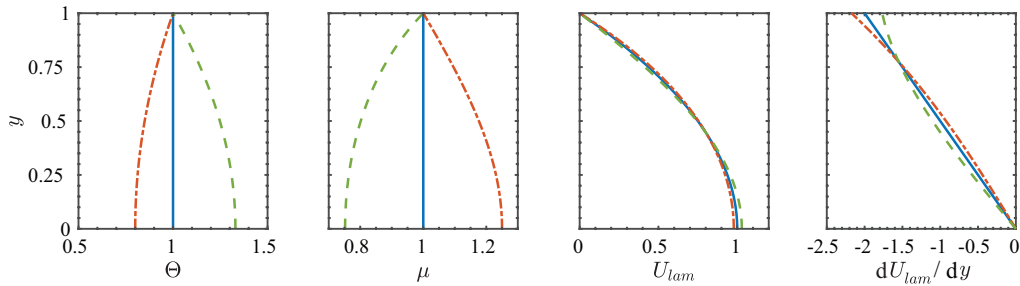}
\caption{Exemplary temperature distributions and resulting viscosity, velocity and velocity gradient profiles. Due to symmetry (only $\mathrm{d}U_{lam} / \mathrm{d}y$ is anti-symmetric), only the upper half channel is shown. The wall is located at $y=1$ and the centerline at $y=0$. Lines correspond to $\mu=\textrm{const}$, or $\mu_c=1$ (solid); $\mu_c=0.75$ (dash); $\mu_c=1.25$ (dash-dot). Line patterns are consistent in the four panels.}
\label{fig:base_example}
\end{figure}

Equation~(\ref{eq:visc}) allows to integrate analytically the streamwise momentum equation and get the following non-dimensional laminar velocity
\begin{equation}\label{eq:vlam}
U_{lam}(y) = -\frac{10}{4\Theta_c+6} \left( \Theta_c y^2 + \frac{1-\Theta_c}{2}y^4 - \frac{1+\Theta_c}{2} \right),
\end{equation}
which was made non-dimensional using the centerline velocity at constant temperature $U_c=-\frac{1}{2}\frac{\mathrm{d}P}{\mathrm{d}x}h^2$ and where the factor $(4\Theta_c+6)/15$ was used to transform bulk to centerline units.
Figure~\ref{fig:base_example} shows some exemplary temperature, viscosity, laminar velocity and velocity gradient profiles.
The latter is denoted as $\mathrm{d}U_{lam} / \mathrm{d}y$.
Increasing viscosity towards the centerline (dash-dot lines) results in a velocity profile that is fuller and linearly more stable if compared to the one of a fluid with constant viscosity.
This is confirmed by a larger (in magnitude) velocity gradient near the walls and a smaller gradient in the central part of the channel.
The opposite effect results from decreasing viscosity towards the centerline (dash lines).
Note that the changes in viscosity reported in the figure and discussed throughout the paper can be achieved with moderate heating or cooling in practical applications.
For example, an increase or decrease of viscosity of water by $\pm 25\%$ from a reference condition of $\Theta_{ref} = 15^\circ \, \mathrm{C}$ is obtained with a decrease or increase of temperature by $\Delta \Theta = -8.5^\circ \, \mathrm{C}$ or $\Delta \Theta = 12^\circ \, \mathrm{C}$, respectively.
Temperature gradients can be significantly larger in real life problems, and so does the variation in viscosity.

In the proceeding of the paper, flow cases with different viscosity distributions, which will be later identified by the centerline value $\mu_c$ only, will be compared at constant Reynolds number based on i) wall viscosity ($\overbar{Re}$); ii) average viscosity across the channel height
\begin{equation}\label{eq:re}
Re = \frac{2}{\int_{-1}^{1}\mu \, \mathrm{d}y}\frac{\rho^*_w h \overbar{U}_c}{\mu^*_w} = \frac{1}{\mu_{av}}\overbar{Re}.
\end{equation}
For a constant viscosity flow $\overbar{Re}=Re$ and $\overbar{U_c}=U_c$.
Definition~(\ref{eq:re}) allows us to filter out the bulk effect of having locally larger or smaller viscosity with respect to the reference value at the walls~\citep{1996_WallAndWilson, 2007_SameenAndGovindarajan}.
On the other hand, in the context of shear-thinning fluids, it has been shown by~\citet{2007_NouarEtAl} that the correct viscosity scale for the definition of the Reynolds number is the one at the wall. 
Choosing a different reference value might yield qualitatively wrong conclusions on the effect of viscosity on the stability of the flow.
For this reason, we will carry out comparison using both $\overbar{Re}$ and $Re$ in order to confirm that our discussion does not depend on the specific choice of the reference viscosity.
In the following, all quantities are made non-dimensional using the semi-channel height, $h$, and the centerline velocity of the flow at constant viscosity, $U_c$.
The ratio between centerline velocities for variable and constant viscosity cases when comparing at constant $Re$ is given by $\overbar{U}_c/U_c = \overbar{Re}/Re = \int \mu \, \mathrm{d}y /2$.
We introduce the following definitions of volume averaged perturbation kinetic energy (PKE) based on the streamwise and cross-flow components of the velocity field, $E_{sw}$ and $E_{cf}$, that will be used in the discussion of the results
\begin{eqnarray}\label{eq:energy_def}
E_{sw} & = & \frac{1}{2}\frac{1}{L_x L_z}\int_{0}^{L_z}\int_{0}^{L_x}\int_{-1}^1 \left( u(x,y,z) - u_{mean}(y)\right)^2 \mathrm{d}y \, \mathrm{d}x \, \mathrm{d}z,\\
E_{cf} & = & \frac{1}{2}\frac{1}{L_x L_z}\int_{0}^{L_z}\int_{0}^{L_x}\int_{-1}^1 v^2(x,y,z) + w^2(x,y,z) \,\mathrm{d}y \, \mathrm{d}x \, \mathrm{d}z.
\end{eqnarray}
The total perturbation energy is $E_{tot} = E_{sw} + E_{cf}$.
Additionally, we define volume averaged vorticity
\begin{equation}
\Omega_i = \frac{1}{2}\frac{1}{L_x L_z}\int_{0}^{L_z}\int_{0}^{L_x}\int_{-1}^1 |\omega_i| \mathrm{d}y \, \mathrm{d}x \, \mathrm{d}z.
\end{equation}

\subsection{Validity of the frozen temperature profile assumption}
\label{sec:froz_temp}

We assess the validity of assuming a frozen temperature profile by means of \emph{a priori} and \emph{a posteriori} arguments. 
This assumption corresponds to entirely neglecting the interaction between the fluctuating velocity and temperature, hence viscosity, fields.
Depending on the flow case under consideration, this can be a crude approximation that has been shown to yield wrong conclusions in the context of the linear stability of shear-thinning fluids~\citep{2007_NouarEtAl}.

\begin{figure}
\centering
\subfigure[]{\includegraphics[height=0.35\textwidth]{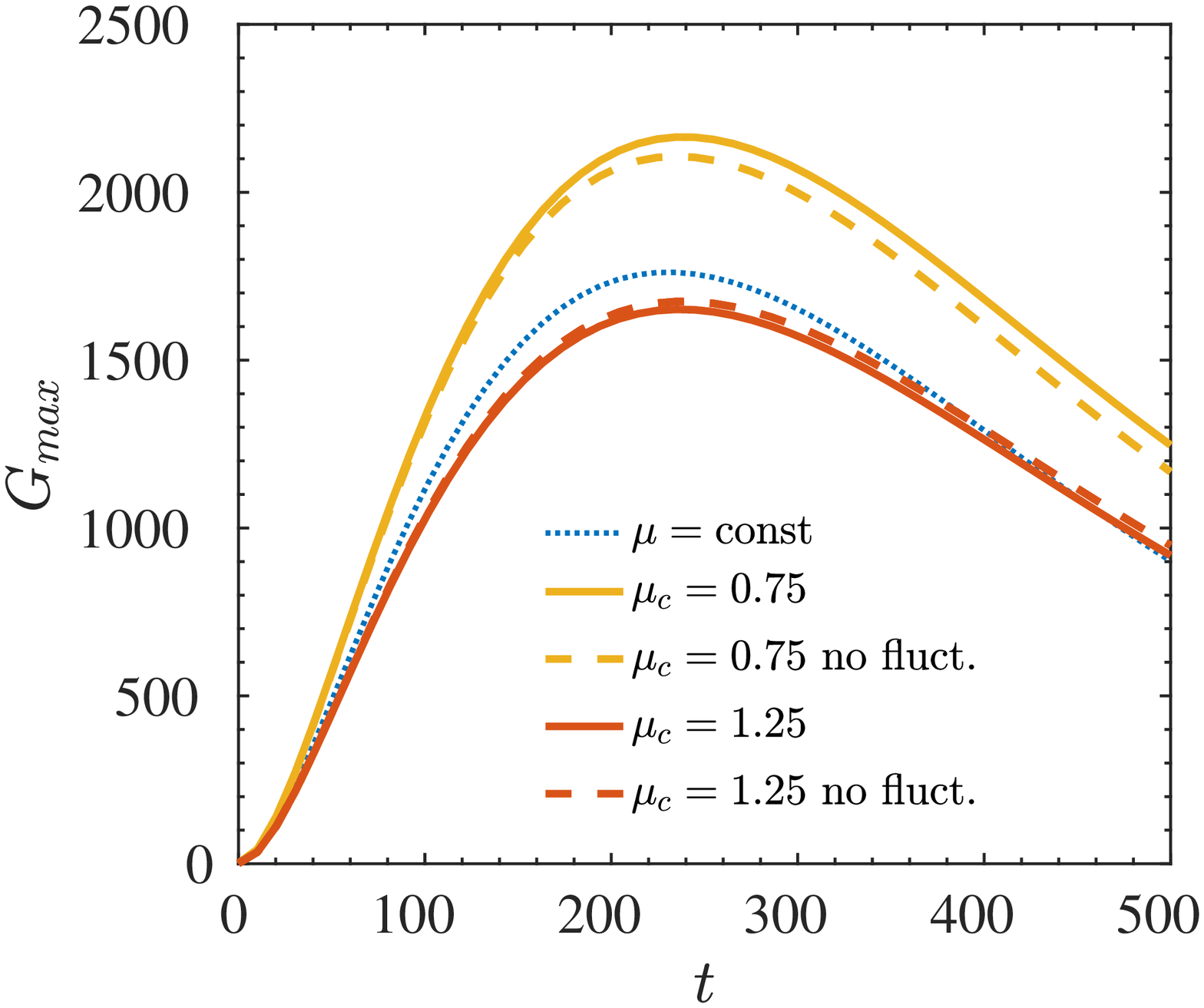}\label{fig:G_fluc_Re}}
\subfigure[]{\includegraphics[height=0.35\textwidth]{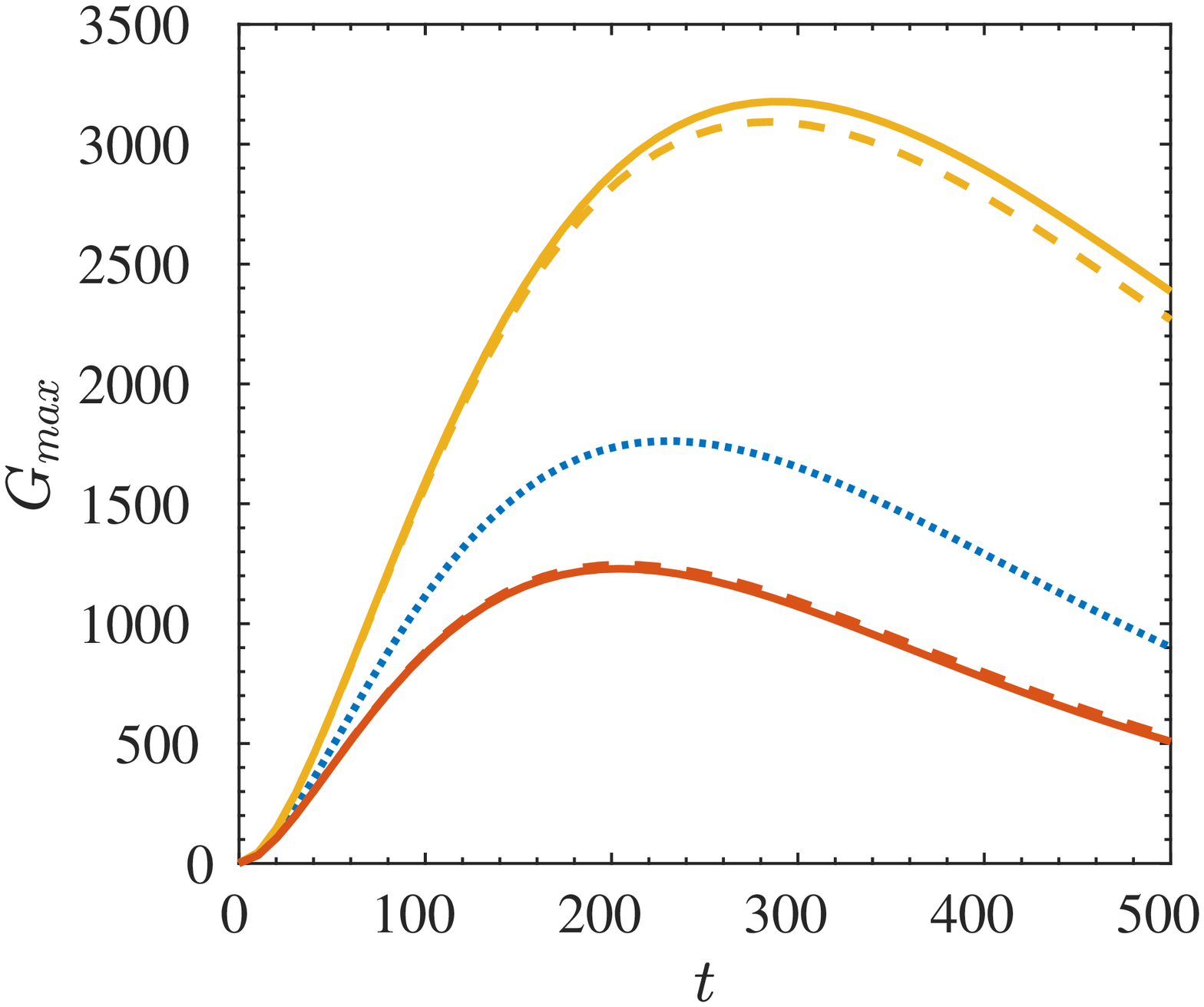}\label{fig:G_fluc_Rebar}}
\caption{Time evolution of transient energy growth function $G_{max}$ for $Pr=7$, $\beta=2$, $Re=3000$ (left) and $\overbar{Re}=3000$ (right).}
\label{fig:G_fluc}
\end{figure}

The \emph{a priori} check is performed using the linearised Navier--Stokes and energy equations for a fluid with temperature dependent viscosity.
For the sake of brevity, the equations are not reported here and the reader is referred to~\citet{1996_WallAndWilson,2007_SameenAndGovindarajan} for details.
Similarly to~\citet{2007_NouarEtAl}, we have calculated the energy growth function, $G_{max}$, which represents the maximum possible linear amplification in time of small initial perturbations~\citep{1993_ReddyAndHenningson}, and compared the results obtained including and neglecting the temperature fluctuation terms in the momentum equations.
The laminar velocity and temperature profiles used for the linearisation are the ones of equations~(\ref{eq:vlam}) and ~(\ref{eq:temp}).
Figure~\ref{fig:G_fluc} displays the calculated growth functions for Reynolds $Re=3000$ and $\overbar{Re}=3000$, Prandtl number $Pr = \mu / \rho \kappa = 7$ (water), where $\kappa$ is the thermal diffusivity of the fluid.
The perturbation is assumed to be streamwise independent and to have a spanwise wavenumber $\beta=2$.
Three viscosity profiles are considered, namely $\mu=\mathrm{const}$, $\mu_c=0.75$ and $\mu_c=1.25$.
Differently to what was found for shear-thinning fluids, results for a fluid with temperature dependent viscosity confirm that the error committed by excluding the temperature fluctuation terms in the momentum equations is negligible.

\begin{figure}
\centering
\includegraphics[height=0.35\textwidth]{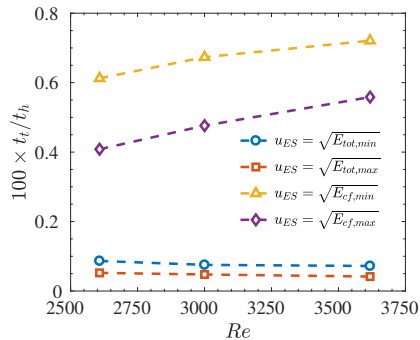}
\caption{Ratio of time scales between the ``turbulent mixing'', $t_t$, and of heat diffusion, $t_h$. The values of the energy are taken from figure~\ref{fig:muc-E} and $Pr=7$.}
\label{fig:timescales}
\end{figure}

The \emph{a posteriori} verification of the validity of the assumption on the temperature profile is based on the estimation of the time scales of the ``turbulent mixing'', $t_t$, and of heat diffusion, $t_h$, namely
\begin{equation}
t_t = \frac{h}{u^*_{ES}}, \qquad t_h = \frac{h^2}{\kappa}, \qquad \frac{t_t}{t_h} = \frac{1}{u_{ES}Pr Re},
\end{equation}
with $u^*_{ES}$ and $u_{ES}$ a characteristic velocity scale of the edge state fluctuating field in dimensional and non-dimensional form, respectively, and $Pr=7$.
The ratio of time scales for three Reynolds numbers, $Re=2608, 3000, 3618$, is shown in figure~\ref{fig:timescales}.
Four reference velocity scales are considered.
They are calculated as the square-root of the maximum and minimum values of the kinetic energy over one period of the edge state.
The used values of $E_{tot}$ and $E_{cf}$ are the ones reported in \S~\ref{sec:variable_visc}, figure~\ref{fig:muc-E}.
Results show a prominent separation of the time scales associated with the fluid dynamic mixing and the diffusion of heat, which supports the assumption that changes in the temperature profile are slow and cannot be seen by the fluctuating velocity field.

As a final note on the strength and role of viscosity fluctuations on the velocity field, we refer to recent studies of heat transfer in turbulent channel flows in the low Mach number limit with strong mean property variations (viscosity and density).
It has been found by~\citet{2015_PatelEtAl,2016_PatelEtAl} that the turbulent viscosity fluctuations are less than $10\%$ of the mean values, even when the variation of the mean viscosity reaches a factor 2, and that the largest effect on turbulent structures and statistics is due to the mean gradients. 

\subsection{Discretisation and edge state identification}

All the simulations are performed using the spectral code SIMSON~\citep{2007_ChevalierEtAl} on a nearly minimal box sized $\pi \, \times \, 2 \, \times \, 0.4\pi$.
The flow is driven by an adaptive pressure gradient that ensures a constant laminar bulk Reynolds number.
The velocity components are expanded in $N_x$ and $N_z$ Fourier modes along the horizontal directions and in $N_y$ Chebyshev polynomials in the wall-normal direction.
A resolution of $N_x \times N_y \times N_z = 48\times97\times48$ was found sufficient to fully resolve the flow for each Reynolds numbers considered.
Dealiasing is performed using the $3/2$ rule.

In order to track edge state solutions we apply a standard bisection algorithm~\citep{2006_SkufcaEtAl} on the amplitude of the initial perturbation velocity field. 
The flow is evolved in time and the integrated root-mean-square (rms) value of the wall-normal velocity, $v_{rms}$, is used to discern whether the flow is laminar or turbulent.
The simulation is stopped if $v_{rms}$ reaches predefined threshold values, typically set to $v_{rms,lam}=2\times10^{-4}$ and $v_{rms,tur}=2.5\times10^{-2}$, respectively.
The tolerance on the scaling coefficient for the initial perturbation velocity field is set to $2\times 10^{-14}$.
In order to remain on the manifold and avoid departures due to numerical errors, the bisection step is repeated at a constant interval $\Delta t=500$.
As a consequence, the relative difference between the energies of the trajectories that relaminarise and become turbulent changes at each restart.
We have verified that relative values remain small for all simulations.
Typically, at restart $\Delta E_{cf} = (E_{cf,tur}-E_{cf,lam}) / E_{cf,lam} = 10^{-5}$; occasional peak values reach $\Delta E_{cf} = 10^{-3}$.
For each combination of Reynolds numbers and centerline viscosity the edge state is tracked in time for at least $30\,000$ non-dimensional time units, unless earlier convergence to a relative periodic orbit is achieved.

\section{Edge states in minimal channels with constant viscosity}
\label{sec:constant_visc}

We start the discussion of the results by introducing the main features of edge states solutions in minimal boxes for canonical channel flows with constant viscosity. 
For such flow configuration edge states break the vertical symmetry of the channel and localise close to one of the two walls, while extending in the streamwise and spanwise directions~\cite{2012_XiAndGraham, 2014_ZammertAndEckhardt_FDR}.
Our main reference in validating the results for $\mu_c=1$ is~\cite{2014_ZammertAndEckhardt_FDR}, who performed edge tracking on the same box size and in the same range of Reynolds numbers.
The typical time evolution of an edge state trajectory is shown in figure~\ref{fig:timeplot} in terms of cross-flow and streamwise energy (figure~\ref{fig:tEn}) and streamwise and wall-normal vorticity (figure~\ref{fig:tOm}).
Bursting events occurring at $t=t_A$ are responsible for generating quasi-streamwise vortices that produce low and high speed streaks by the lift-up mechanism. 
The streaks reach a maximum PKE at the subsequent time $t=t_B$, when the cross-flow motion has been dissipated.
It is possible to look at this two step process in terms of vorticity.
The bursting events correspond to peaks in the volume averaged streamwise vorticity $\Omega_x$, which is a measure of the strength of the quasi-streamwise vortices.
The peaks in streaks intensity as measured by $E_{sw}$ correspond to peaks in wall-normal vorticity $\Omega_y$, which reduces to the shear $\partial u / \partial z$ when the streaks are streamwise independent.
Snapshots of the bursting event at $t_A$ and of the maximum amplitude streaks at $t_B$ are displayed in figure~\ref{fig:snapshots}.
The final step that makes the edge state dynamics self-sustained is an instability of the two-dimensional streaky profile of figure~\ref{fig:snap_streaks} that results in a new burst of the streaks.

Based on the features discussed in figure~\ref{fig:timeplot} and~\ref{fig:snapshots}, we classify edge state orbits into periodic and aperiodic ones using as measure of the periodicity the time between two consecutive maximum peaks in the cross-flow kinetic energy, which will be indicated as $T$.
In the following we will use the terms aperiodic and chaotic interchangeably when discussing the nature of the edge state orbit.

\begin{figure}
\centering
\includegraphics[width=0.6\textwidth]{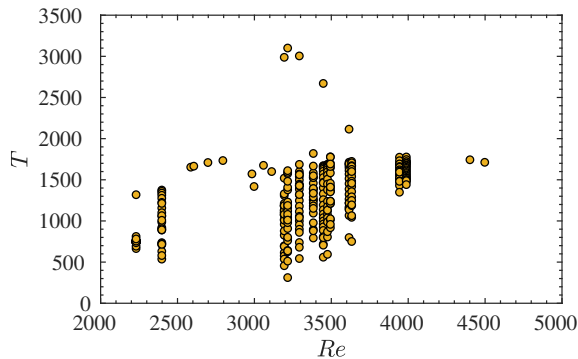}
\caption{Bifurcation diagram of the inter-burst period $T$ as function of $Re$ for channel flows with constant viscosity, $\mu_c=1$. Multiple symbols at a given $Re$ indicate that the state is aperiodic.}
\label{fig:Re-T}
\end{figure}

\begin{figure}
\centering
\includegraphics[width=0.6\textwidth]{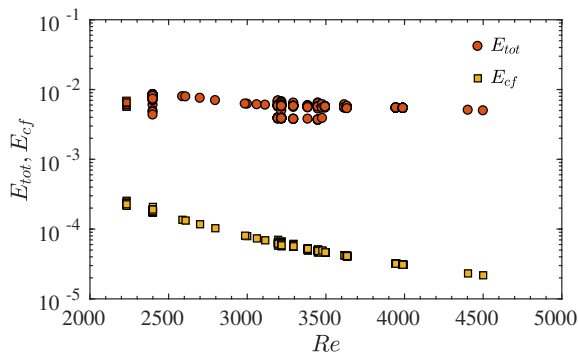}
\caption{Bifurcation diagram of the maximum total and cross-flow energy, $E_{tot}$ and $E_{cf}$, over the edge state evolution for channel flows with constant viscosity, $\mu_c=1$. Multiple symbols at a given $Re$ indicate that the state is aperiodic.}
\label{fig:Re-Etot-Ecf_cv}
\end{figure}

The bifurcation diagram of inter-burst time intervals $T$ over a typical range of Reynolds numbers at which subcritical transition occurs is presented in figure~\ref{fig:Re-T}.
Periods are indicated by symbols at given $Re$; periodic orbits correspond to single entries while for chaotic ones we report all the calculated periods over the total integration time after discarding the initial transient up to $t=4000$.
Results are in agreement with~\citet{2014_ZammertAndEckhardt_FDR} and show mostly periodic solutions in the range $Re=[2600,3100]$ and for $Re>4000$.
At intermediate Reynolds, $Re=[3100,3600]$, orbits undergo bifurcations that result in chaotic states, for which $T$ largely fluctuates.
In the range $Re=[3100,3300]$, we could not track the $PO_2$ family of periodic orbits documented by~\citet{2014_ZammertAndEckhardt_FDR} characterised by short periods $T\approx500$.
On the contrary, we find chaotic recurrence of bursts even after extending the total tracking time of the states to twice the one used by the mentioned authors.
The chaotic nature of the edge state recurrence depends on the properties of the dynamic saddle and does not bear a specific physical meaning. Minute changes to any flow parameter can affect significantly the recurrence of the edge state, as it is the case for the relative periodic orbit documented by~\citet{2013_KhapkoEtAl} in the asymptotic suction boundary layer. Small changes resulted in a spanwise left-hopping, right-hopping, left-right-hopping or erratic shifts. Despite this substantial difference, the nature of the flow structures evolving in time remains the same between the four cases. Similarly, the edge states reported in this paper are qualitatively the same as in figures~\ref{fig:timeplot} and~\ref{fig:snapshots} for each Reynolds number considered, despite differences in their period. The chaotic orbits of figure~\ref{fig:Re-T} and the ones described in \S~\ref{sec:variable_visc} can also be explained by the fixed maximum observation time used in the simulations; it cannot be excluded that extending this limit to sufficiently large times could result in convergence to a constant time period.

The bifurcation diagrams of the maximum total and cross-flow energy during each recurrence of the edge state regeneration cycle are shown in figure~\ref{fig:Re-Etot-Ecf_cv}.
They provide an indication of the strength of the flow structures and on the minimal energy needed to have self-sustained dynamics. They also represent a local threshold for transition for perturbations evolving in the neighbourhood of the edge state.
There, the slightest deviation from the reported values will either result in relaminarisation or in transition to the turbulent regime.
Overall, both $E_{tot}$ and $E_{cf}$ decrease as the Reynolds number increases, meaning that smaller perturbations measured in centerline velocity are needed to sustain the edge state dynamics.
Several aperiodic edge states are characterised by approximately constant values of $E_{tot}$ and $E_{cf}$ at each regeneration cycle. For example, at $Re=3946$ the period fluctuates between $T=1340$ and $T=1770$ while the energies vary by less than $0.2\%$ around the values $E_{tot}=5.45\times10^{-3}$ and $E_{cf}=3.19\times10^{-5}$.
The physical interpretation of having same perturbation energy and different periods is that, while the volume average energy of the streaks and vortices is repeatedly the same, the specific shape of the streaks changes, thereby modifying their secondary stability and characteristic time scales of the sinuous instability.

\section{Effect of viscosity}
\label{sec:variable_visc}

\subsection{Energy level and local threshold of the edge} 
\label{sec:energy}
In this section we quantify the modification to the perturbation kinetic energy of the edge state caused by variable viscosity.
As discussed in \S~\ref{sec:intro}, an altered energy level of the edge state means a shift of its location in the state space with respect to the laminar attractor and turbulent saddle.
Lower energy corresponds to an edge state that is closer to the laminar state and to a modulation of the surrounding manifold that restricts its basin of attraction.
As a consequence, perturbations evolving near the attracting region of the edge state have to exceed a lower energy threshold to evolve into turbulence.
The opposite results if the energy of the edge state increases.

\begin{figure}
\centering
\subfigure[]{\includegraphics[height=0.33\textwidth]{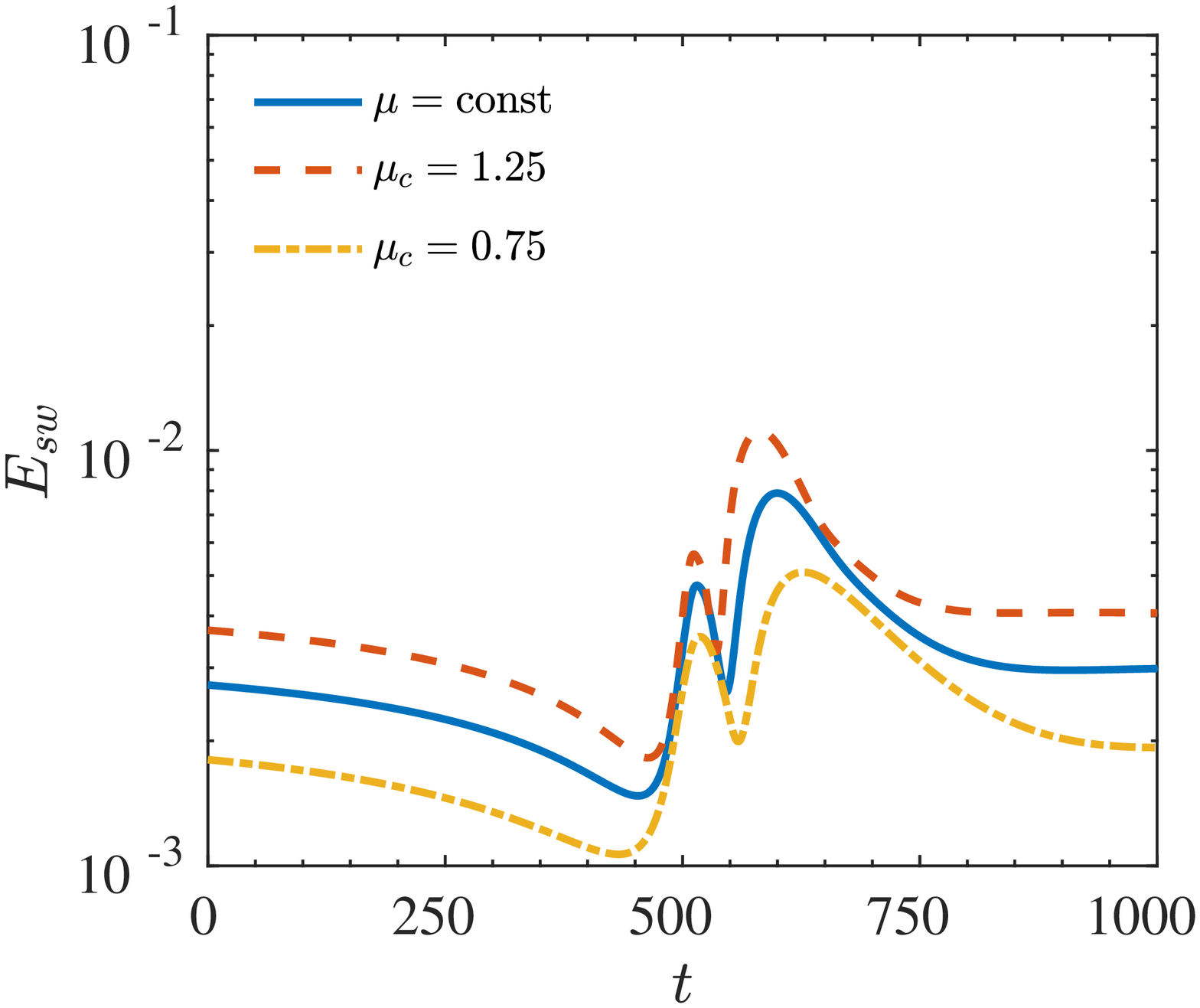}\label{fig:ensw}}
\subfigure[]{\includegraphics[height=0.33\textwidth]{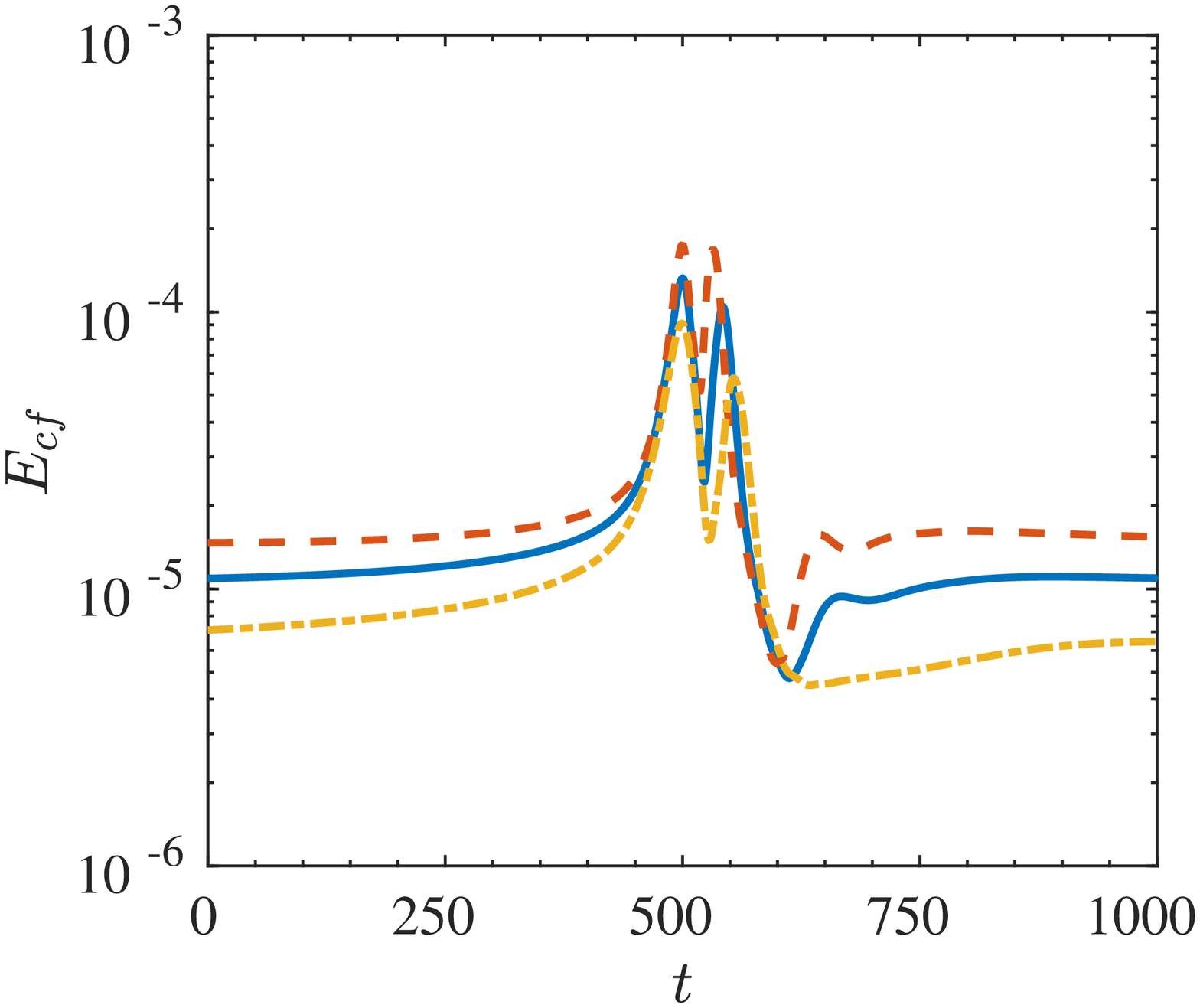}\label{fig:encf}}
\caption{Portion of the time evolution of the volume averaged streamwise and cross-flow kinetic energy, $E_{sw}$ and $E_{cf}$, for flow cases at $Re=2608$. Lines patterns are the same in (a) and (b).}
\label{fig:energy}
\end{figure}

\begin{figure}
\centering
\subfigure[]{\includegraphics[height=0.33\textwidth]{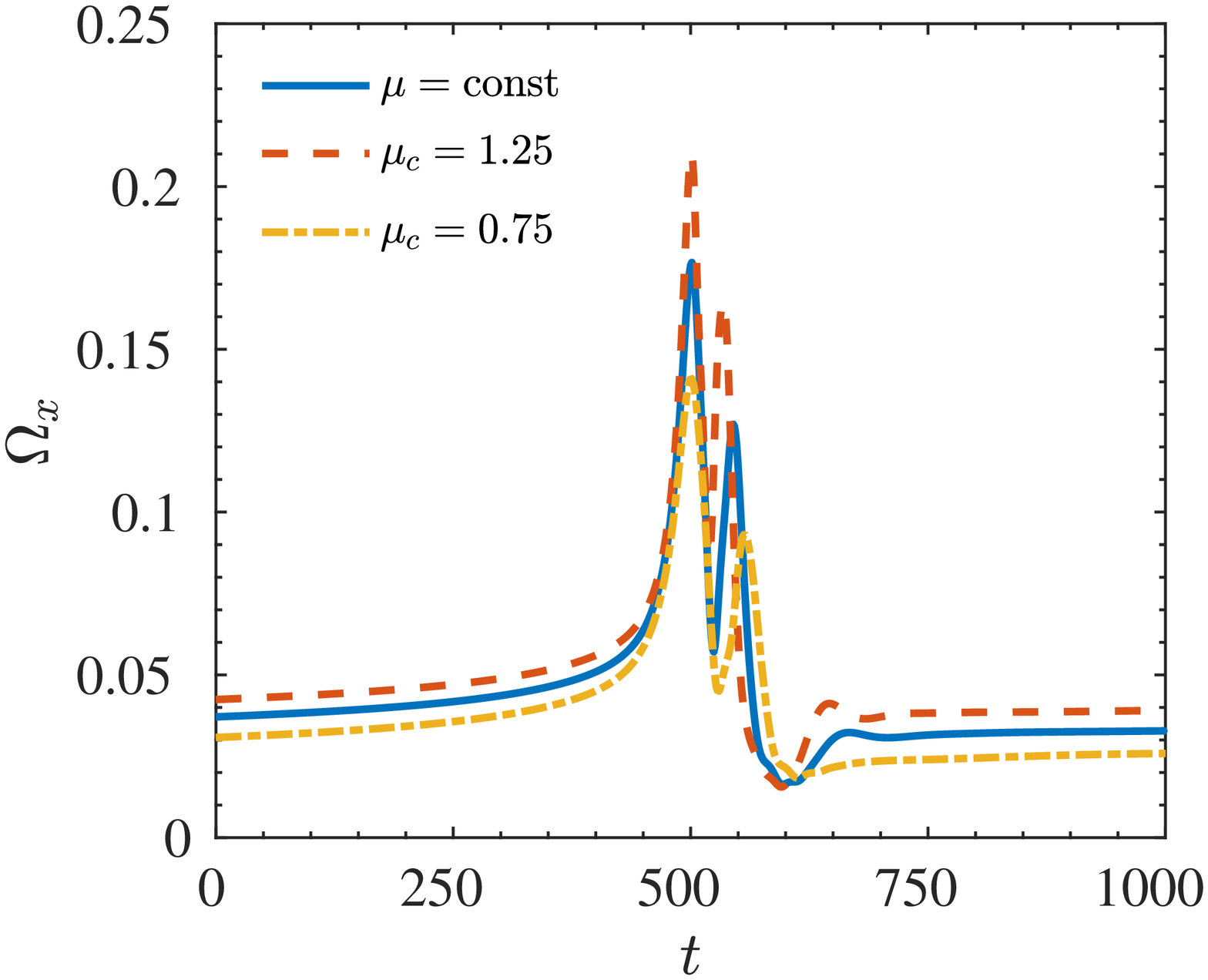}\label{fig:omegax}}
\subfigure[]{\includegraphics[height=0.33\textwidth]{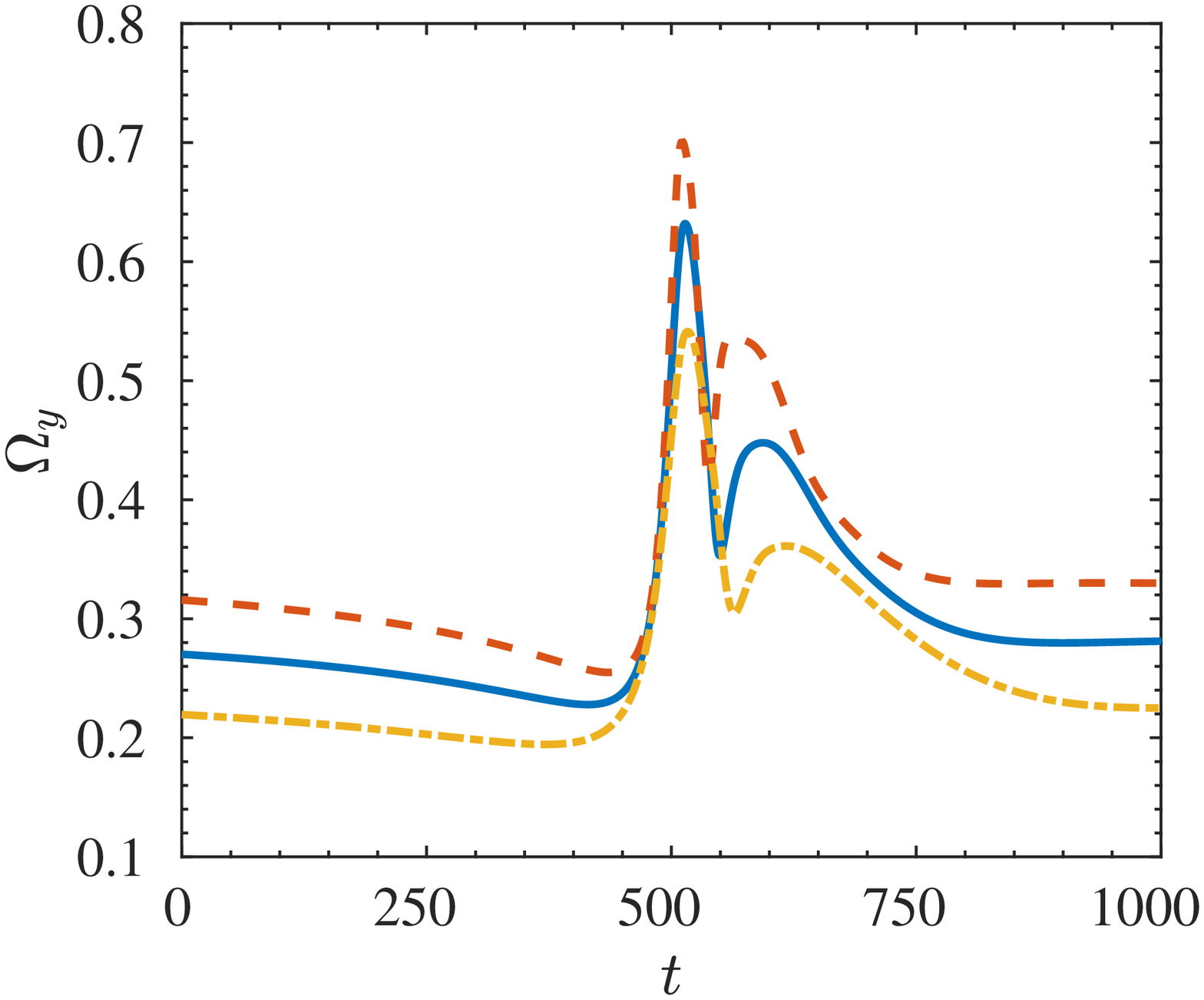}\label{fig:omegay}}
\caption{Portion of the time evolution of the volume averaged streamwise and wall-normal vorticity, $\Omega_x$ and $\Omega_y$, for flow cases at $Re=2608$. Lines patterns are the same in (a) and (b).}
\label{fig:omega}
\end{figure}

\begin{figure}
\centering
\subfigure[]{\includegraphics[height=0.33\textwidth]{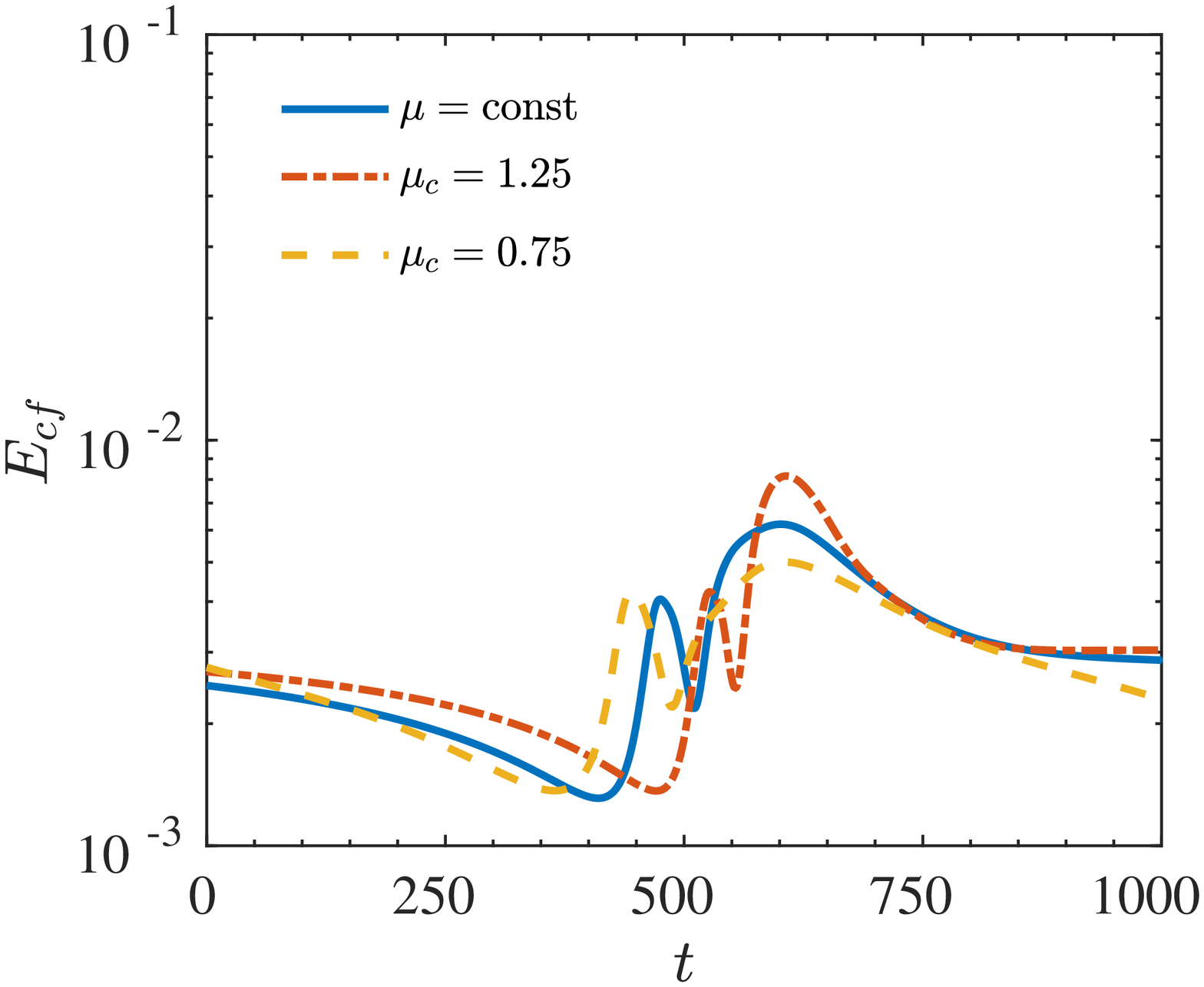}\label{fig:enswRew}}
\subfigure[]{\includegraphics[height=0.33\textwidth]{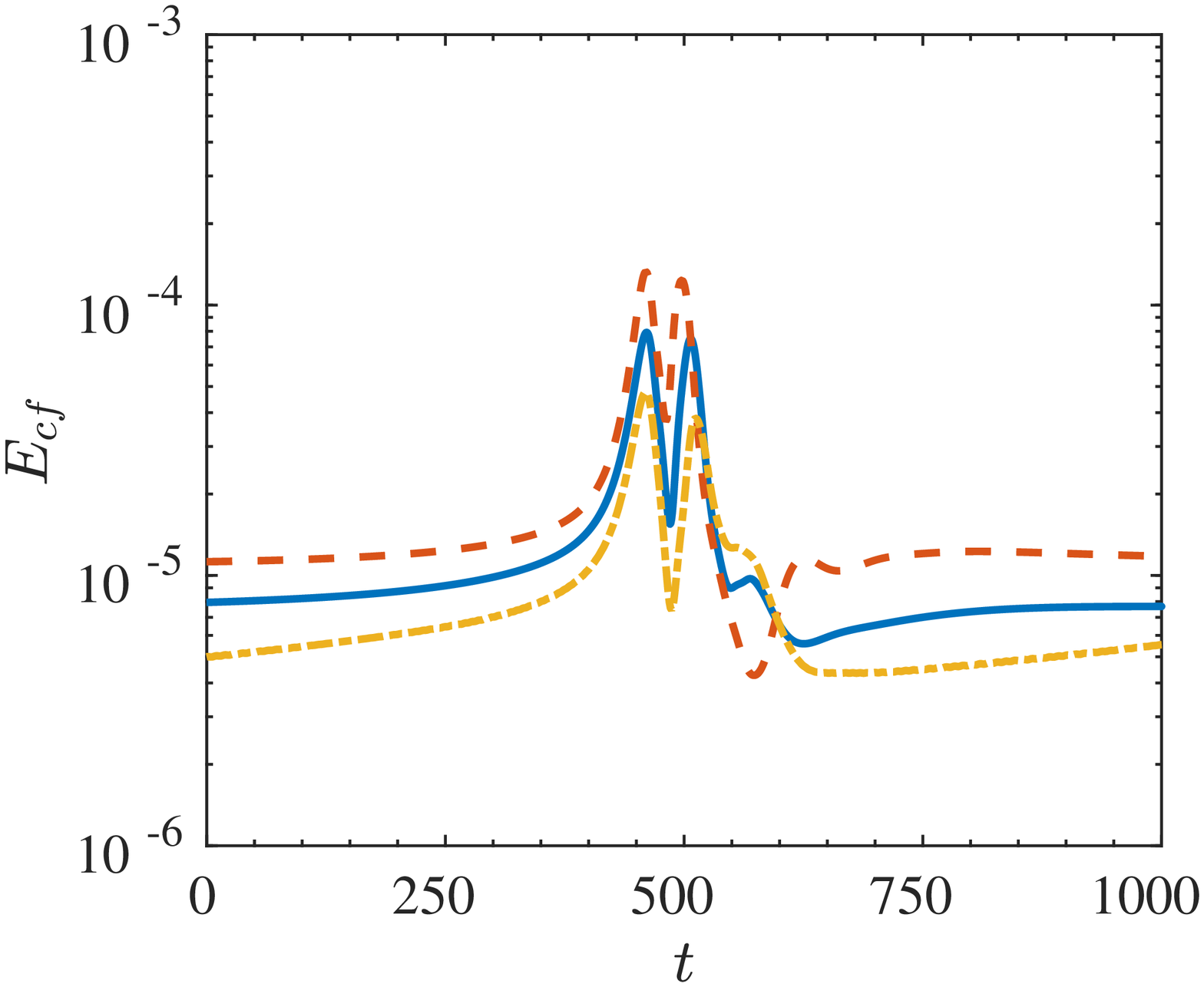}\label{fig:encfRew}}
\caption{Portion of the time evolution of the volume averaged streamwise and cross-flow kinetic energy, $E_{sw}$ and $E_{cf}$, for flow cases at $\overbar{Re}=3000$. Lines patterns are the same in (a) and (b).}
\label{fig:energyRew}
\end{figure}

We start characterising the effect of viscosity by looking at the time evolution of the streamwise and cross-flow energy of the edge states at $Re=2608$ displayed in figure~\ref{fig:energy}.
Variable viscosity does not affect the qualitative trends but modulates the energy levels of the edge state.
If $\mu_c>1$, $E_{sw}$ is larger than for the constant viscosity case at each step of the edge state evolution, while the opposite is true if $\mu_c<1$, see figure~\ref{fig:ensw}.
The same behaviour is followed by $E_{cf}$ in figure~\ref{fig:encf}.
Weaker streaks and weaker vortices are needed in case $\mu_c<1$ to self-sustain the edge state dynamics, therefore the perturbation energy threshold for transition in that area of the state space is smaller.
This statement can also be assessed in terms of streamwise and wall-normal vorticity.
Figure~\ref{fig:omegax} shows that $\Omega_x$ increases for the $\mu_c=1.25$ case with respect to $\mu_c=1$, thus confirming stronger streamwise vortical structures; $\Omega_y$ also increases, see figure~\ref{fig:omegax}, therefore the shear due to streaky structures at their highest amplitude is larger.
The same conclusions are drawn if results are compared at the same $\overbar{Re}$.
Figure~\ref{fig:energyRew} shows $E_{sw}$ and $E_{cf}$ for flow cases with $\mu_c=0.75, \, 1 \, \mathrm{and} \, 1.25$ at $\overbar{Re}=3000$.

\begin{figure}
\centering
\subfigure[]{\includegraphics[height=0.35\textwidth]{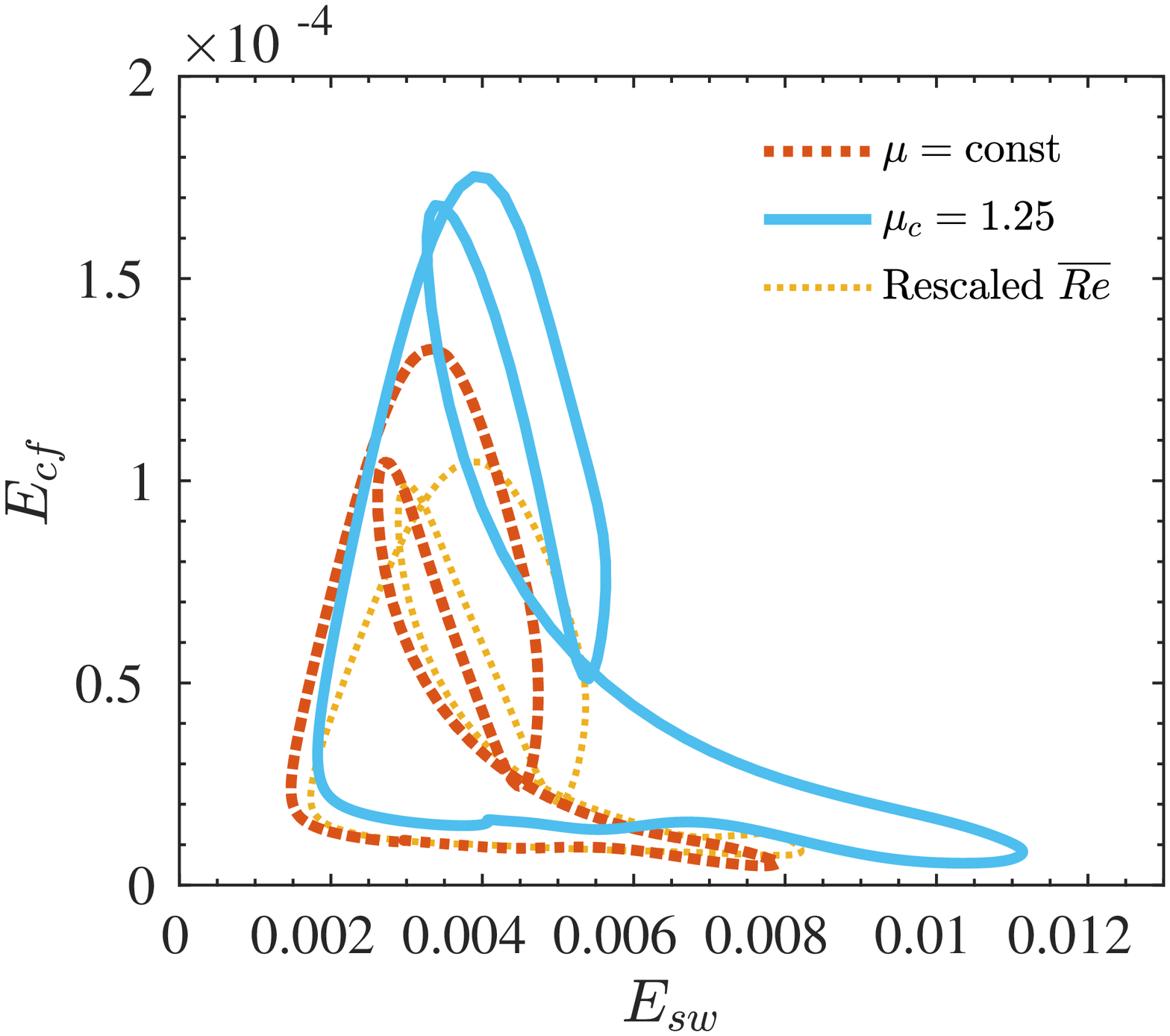}\label{fig:orbit_T080}}
\subfigure[]{\includegraphics[height=0.35\textwidth]{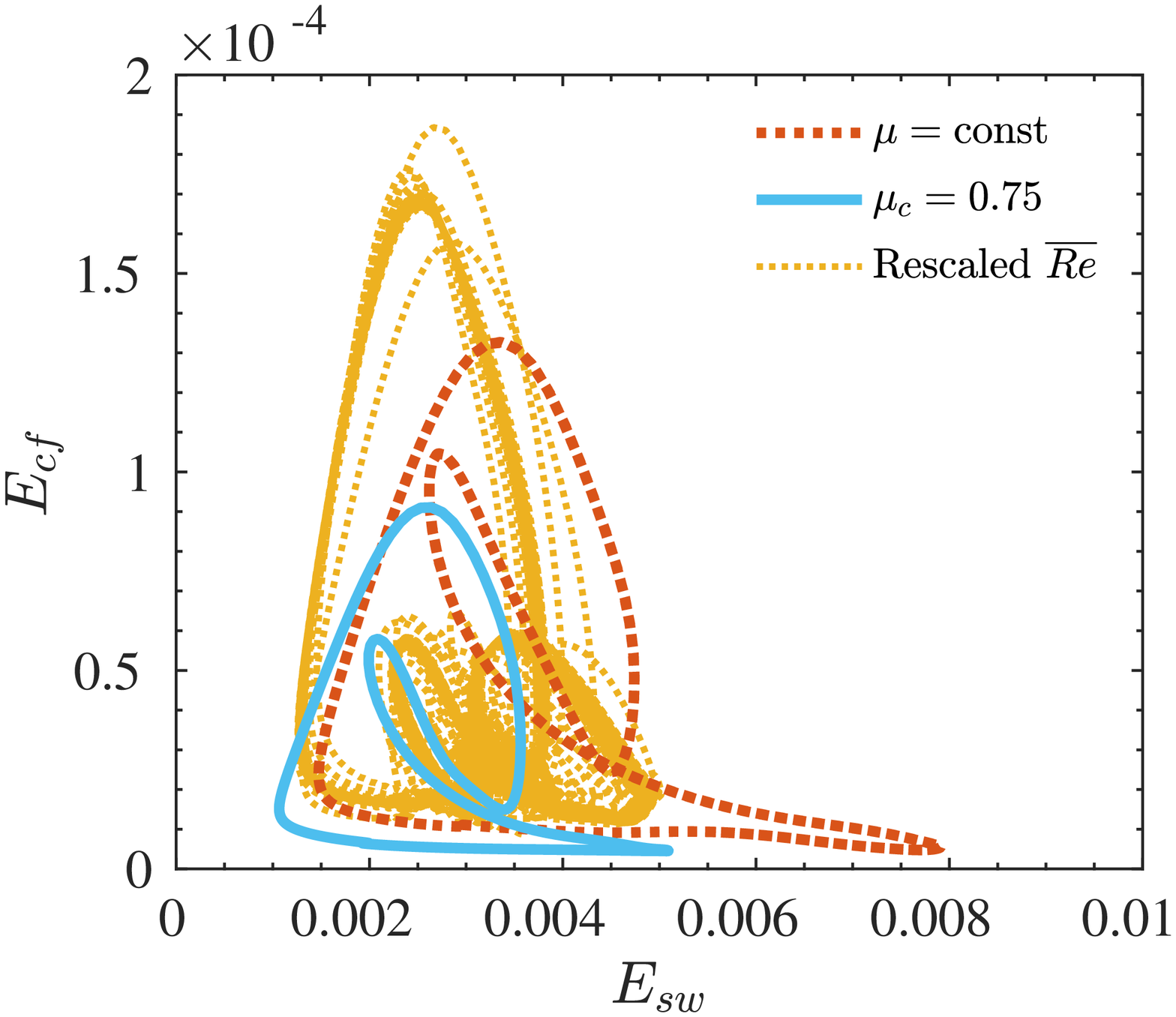}\label{fig:orbit_T133}}
\caption{State space projection of the edge state orbit at $Re=2608$. Rescaled simulations correspond to $\overbar{Re}=3000$ in (a) and $\overbar{Re}=2233$ in (b). Scattered trajectories indicate that the state is aperiodic.}
\label{fig:orbit}
\end{figure}

\begin{figure}
\centering
\includegraphics[height=0.35\textwidth]{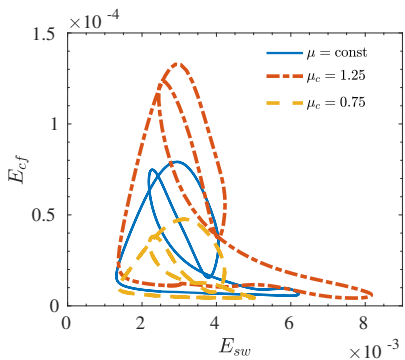}
\caption{State space projection of the edge state orbits at $\overbar{Re}=3000$ and $\mu_c=0.75, \, 1, \, 1.25$.}
\label{fig:orbitRew}
\end{figure}

A complete picture on the whole edge state dynamics modification is given by the projection of its orbit on the $E_{sw}-E_{cf}$ state space, see figure~\ref{fig:orbit}.
Results are presented for $Re=2608$ only but are qualitatively the same at different Reynolds numbers.
The whole integrated history of the energy is plotted after the initial transient of 4000 non-dimensional time units.
When the orbit is periodic the time trace collapses over one single trajectory; this is the case for all the orbits displayed in figure~\ref{fig:orbit_T080}.
On the other hand, aperiodic orbits follow slightly different trajectories over each period thus resulting in a more scattered plot, as is the case for the rescaled $\overbar{Re}=2233$ in figure~\ref{fig:orbit_T133}.
Increasing viscosity away from the walls results in stronger streaks that are generated by stronger vortices as compared to the $\mu_c=1$ case.
It can be then concluded that stronger perturbations are needed to trigger transition in flows evolving near the edge state.
The opposite results from decreasing viscosity towards the centerline.
In order to rule out that the observed effects can be reproduced by a simple rescaling of the Reynolds number of constant viscosity cases when comparing at constant $Re$, the figures additionally include results for $\mu_c=1$ at $\overline{Re} = Re \int \mu(y) \mathrm{d}y / 2$.
Rescaling of constant viscosity simulations, when comparing at constant $Re$, based on the average viscosity, can qualitatively predict the effect of viscosity on the strength of streaks only, while it fails in describing the cross-flow motion and vortex intensity as it predicts an opposite effect.
The same modulating effect of viscosity on the edge state energy is observed comparing flow cases at the same Reynolds number based on the wall viscosity, as displayed in figure~\ref{fig:orbitRew} for $\overbar{Re}=3000$.

\begin{figure}
\centering
\subfigure[]{\includegraphics[height=0.35\textwidth]{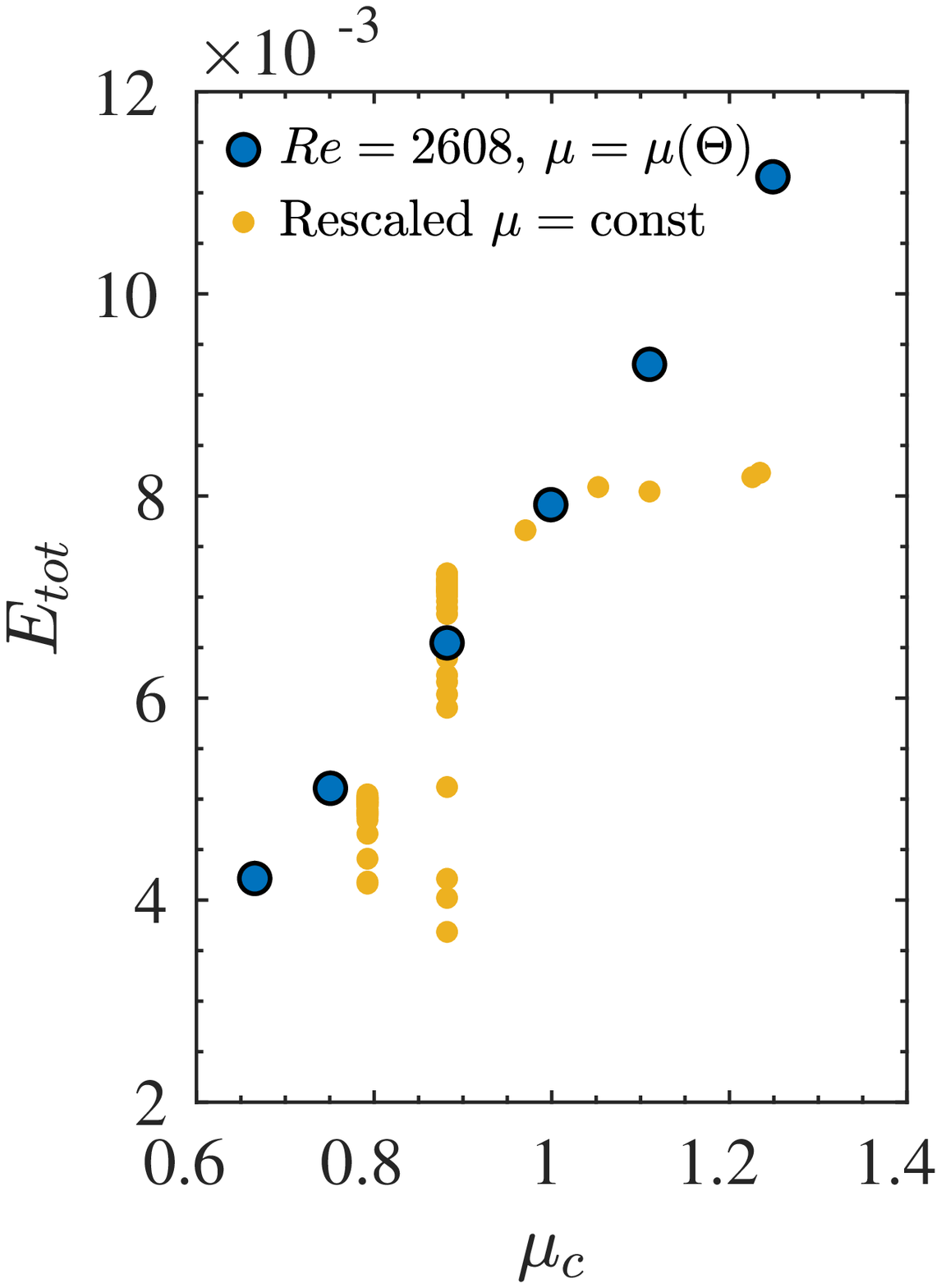}\label{fig:muc-Etot_2608}}
\subfigure[]{\includegraphics[height=0.35\textwidth]{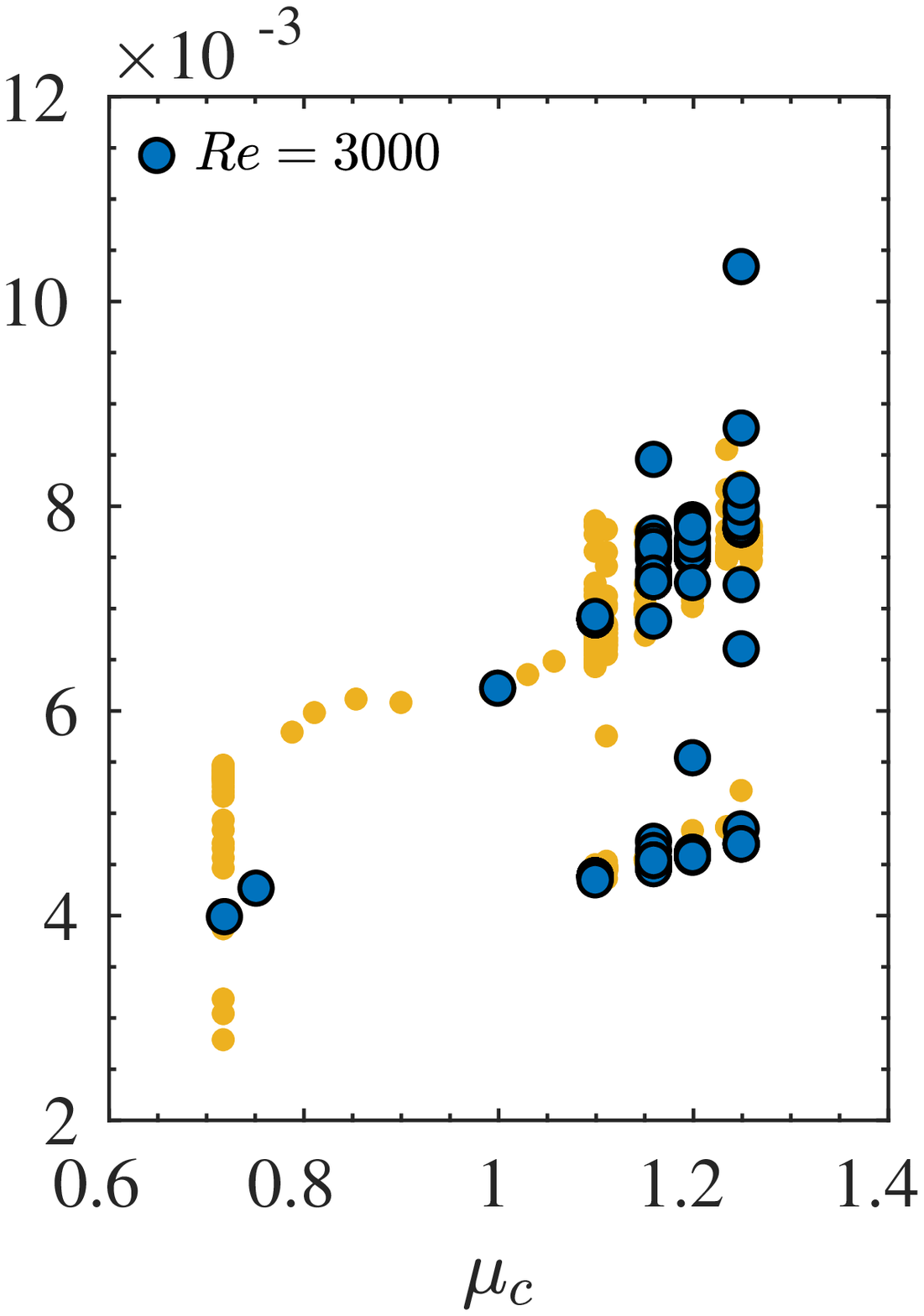}}
\subfigure[]{\includegraphics[height=0.35\textwidth]{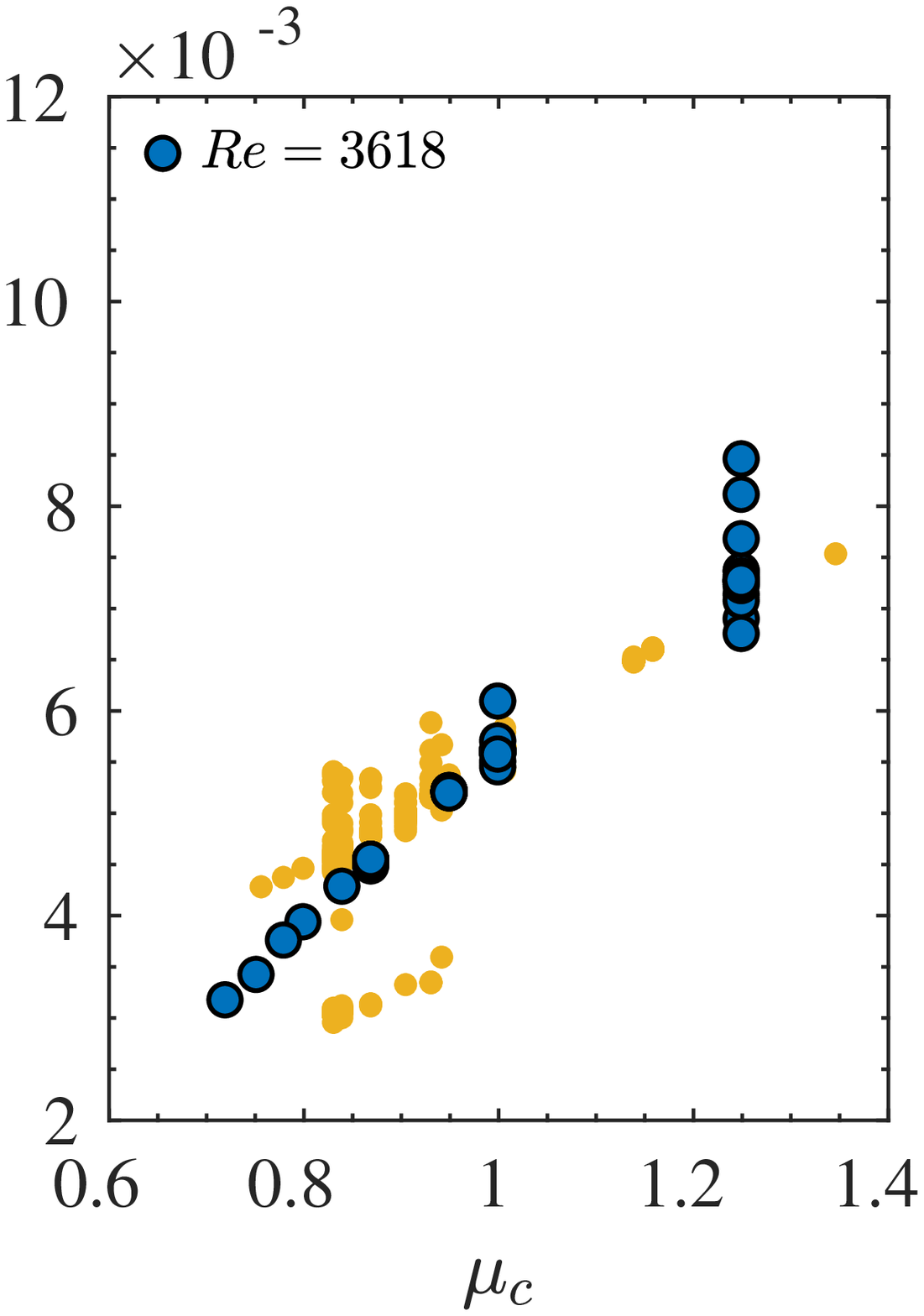}\label{fig:muc-Etot_3618}}\\
\subfigure[]{\includegraphics[height=0.35\textwidth]{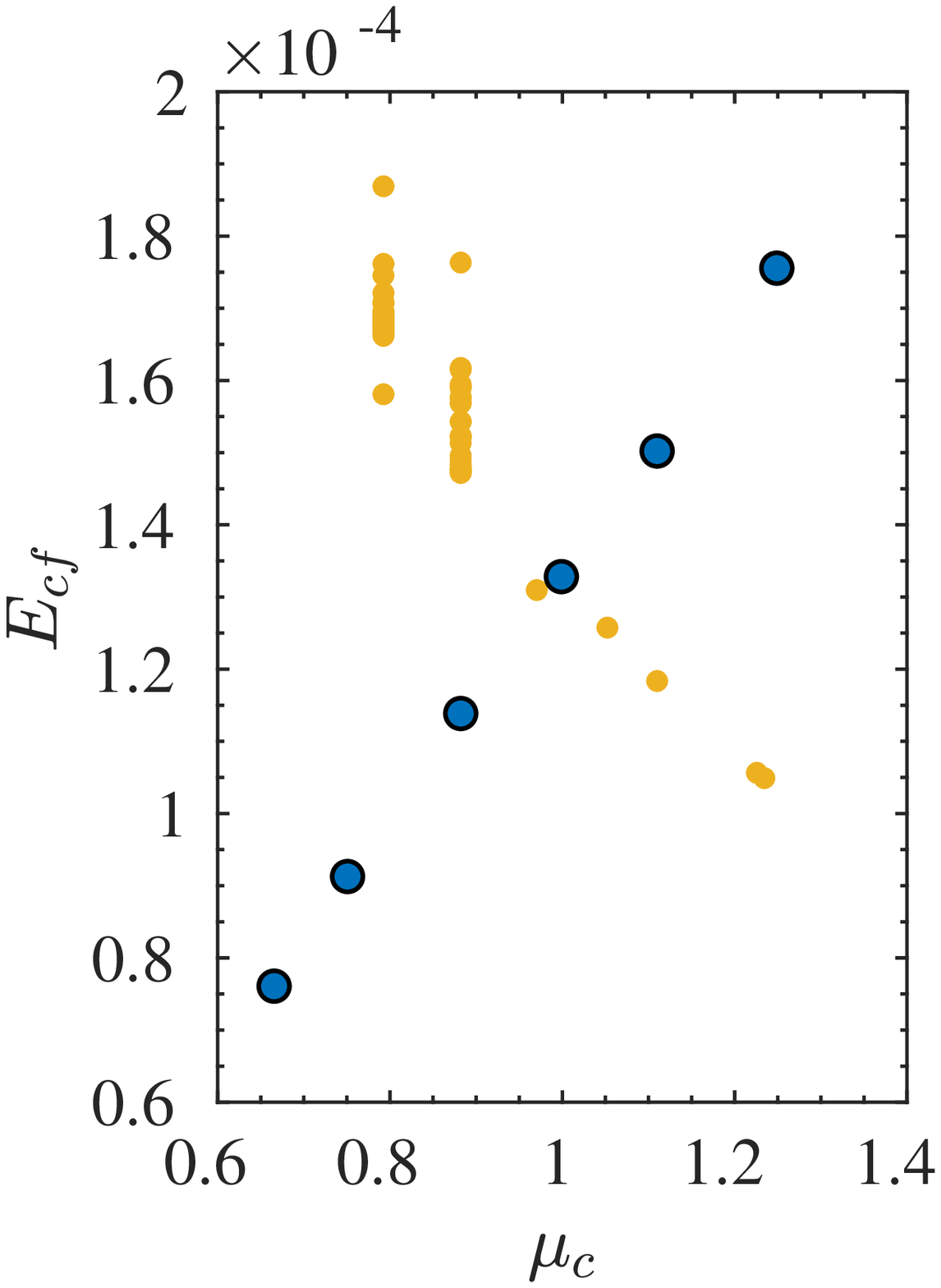}\label{fig:muc-Ecf_2608}}
\subfigure[]{\includegraphics[height=0.35\textwidth]{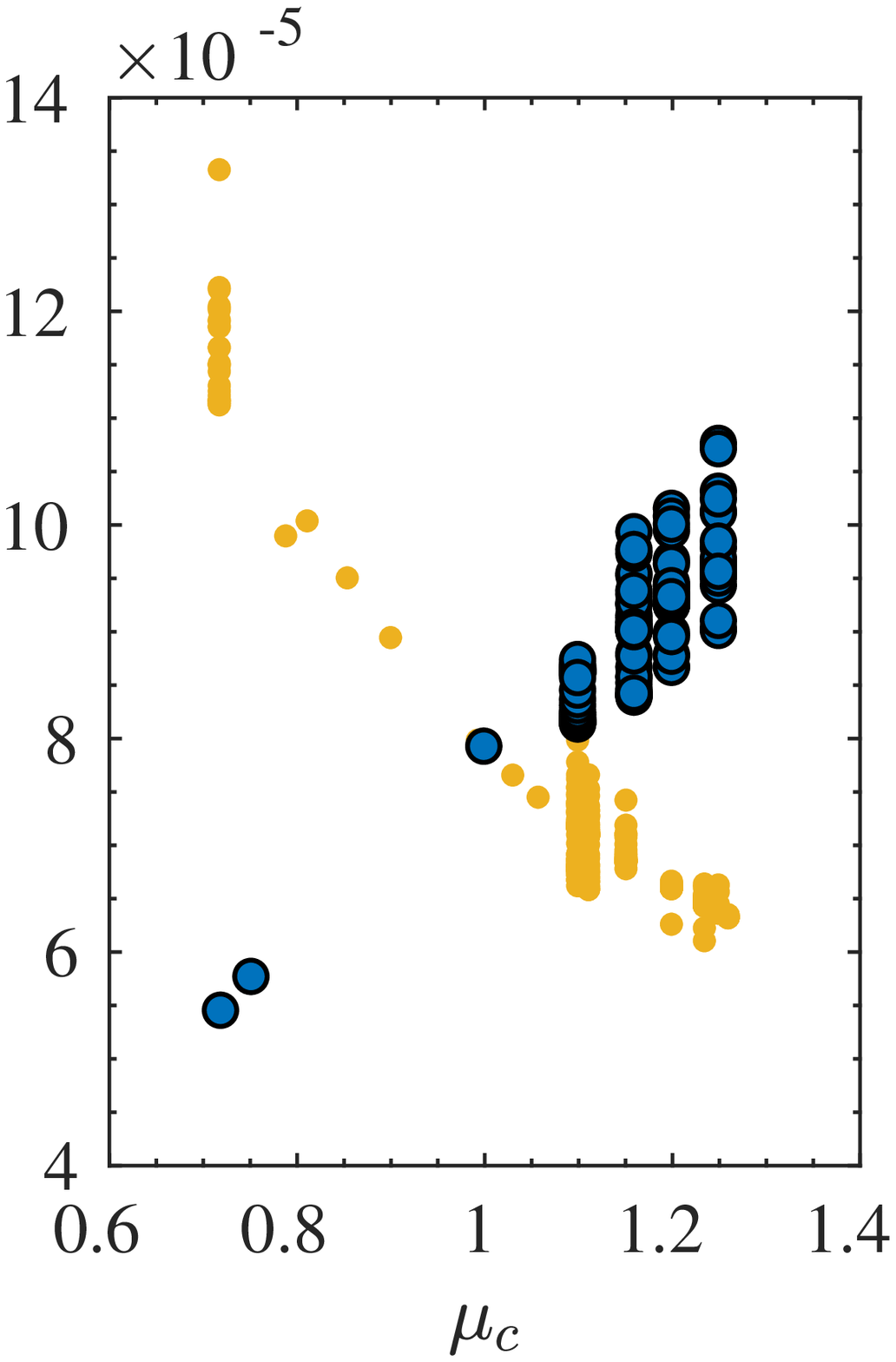}}
\subfigure[]{\includegraphics[height=0.35\textwidth]{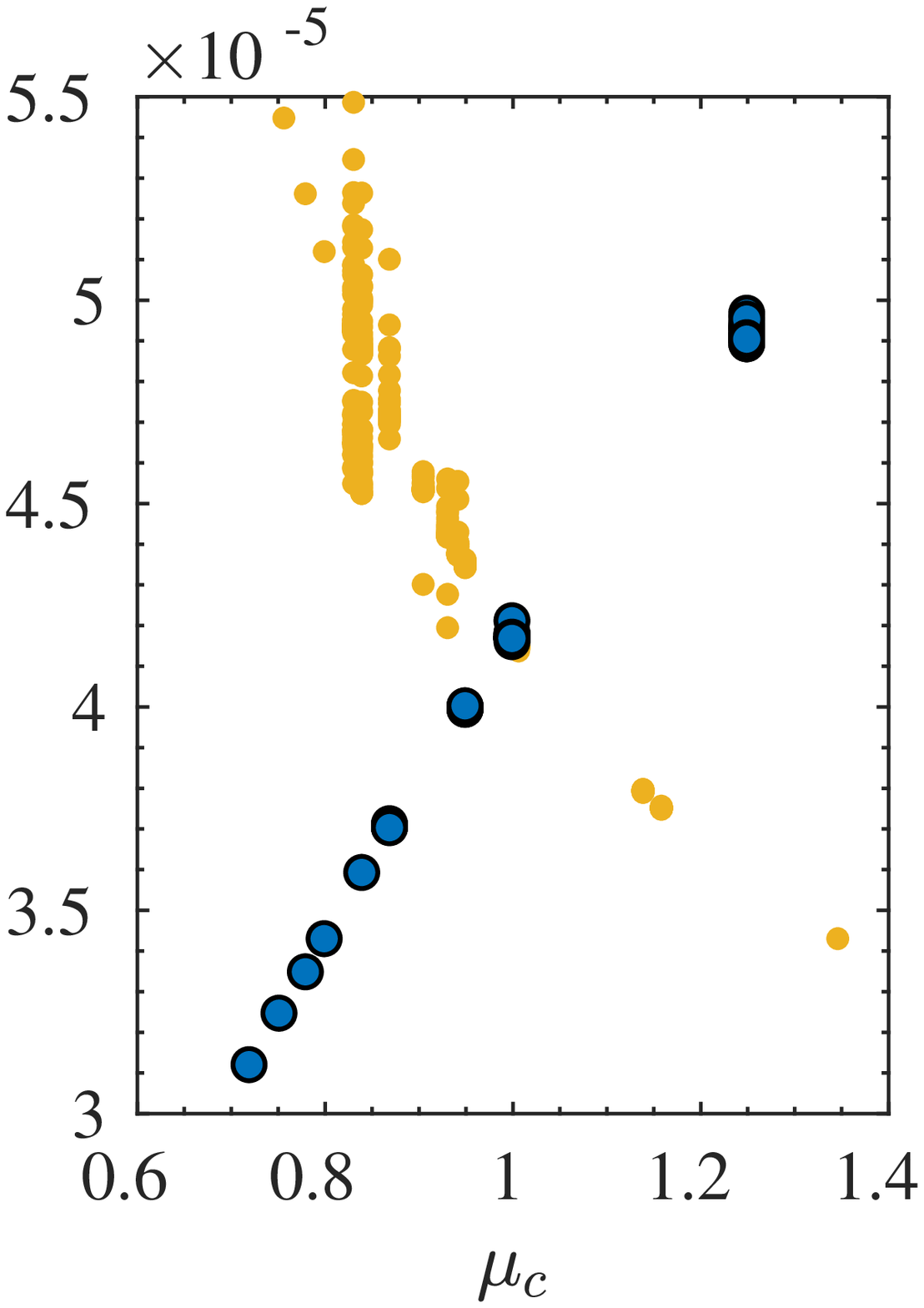}\label{fig:muc-Ecf_3618}}
\caption{Bifurcation diagrams of the maximum (a)-(c) total and (d)-(f) cross-flow energy function of $\mu_c$ for three Reynolds numbers; $Re=2608$ (a) and (d); $Re=3000$ (b) and (e); $Re=3618$ (c) and (f). Symbols are the same as in figure~\ref{fig:muc-T}.  Small light symbols are the edge states for constant viscosity at Reynolds number $\overbar{Re}=Re\, \int \mu \mathrm{d}y \, / 2$; their periods are rescaled by the factor $Re/\overbar{Re}$ in order to be consistent with the normalisation used for the variable viscosity cases.}
\label{fig:muc-E}
\end{figure}

A more comprehensive view on the effect of viscosity on the edge state energy is provided by figure~\ref{fig:muc-E}, which shows at constant $Re$ the bifurcation diagrams of the maximum total and cross-flow energy at each regeneration cycle of the edge state using $\mu_c$ as the bifurcation parameter.
As in figure~\ref{fig:orbit}, results for the constant property cases are included for comparison and rescaled by $\overbar{Re}/Re$.
Smaller viscosity at the centerline consistently results in lower $E_{tot}$ and $E_{cf}$, while the opposite occurs if $\mu_c>1$.
The rescaled values of constant viscosity edge states reasonably capture the overall trend of $E_{tot}$, that is predominantly contained in the streamwise streaks.
However, the quantitative values are not matched and discrepancies are observed in terms of the periodic or chaotic nature of the edge state orbit.
Rescaling of constant viscosity results completely fails in predicting the trend of $E_{cf}$.
If $\mu_c<1$ the maximum kinetic energy of the cross-flow motion monotonically decreases while the opposite occurs in constant viscosity flows at rescaled (higher) Reynolds numbers.
Bifurcation diagrams of maximum streamwise and wall-normal vorticity over the period of the edge state (not included here) confirm what discussed in terms of energy.
The overall increase of maximum $\Omega_x$ if $\mu_c>1$ and decrease if $\mu_c<1$ is consistent with respectively stronger and weaker vortical structures and is opposite to what a rescaling of constant viscosity cases would predict.
The increase ($\mu_c>1$) and decrease ($\mu_c<1$) of the maximum $\Omega_y$ is a measure of the higher and lower shear induced by the strengthening and weakening of the streaks.
Larger peak values of $\Omega_y$ when streaks are strongest and the flow is essentially two-dimensional (the velocity profile is a function of the streamwise and wall-normal co-ordinates only as in figure~\ref{fig:snapshots}) indicate a more pronounced inflectional point from which a stronger secondary instability evolves.

\begin{figure}
\centering
\subfigure[]{\includegraphics[height=0.35\textwidth]{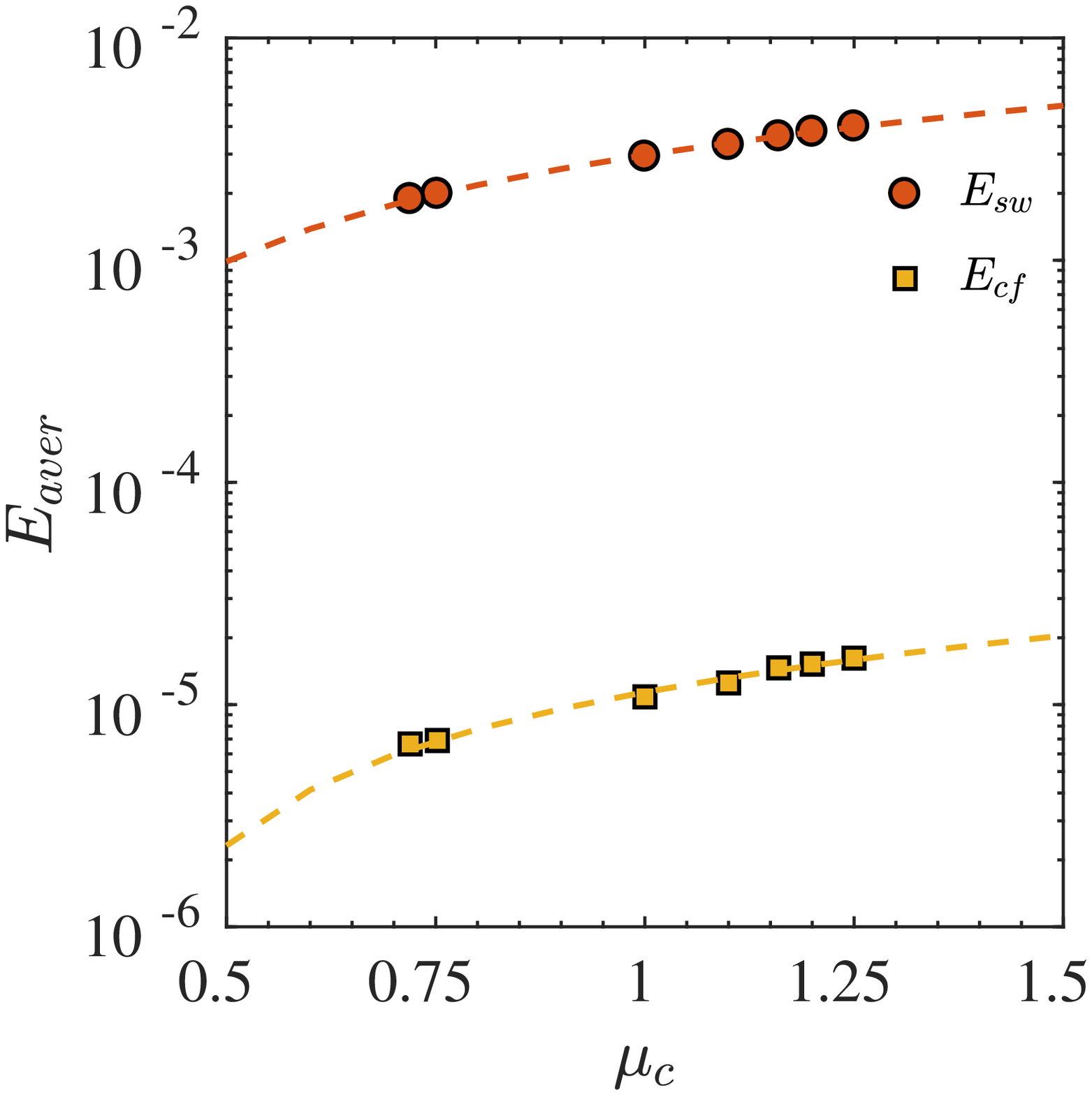}} \quad
\subfigure[]{\includegraphics[height=0.35\textwidth]{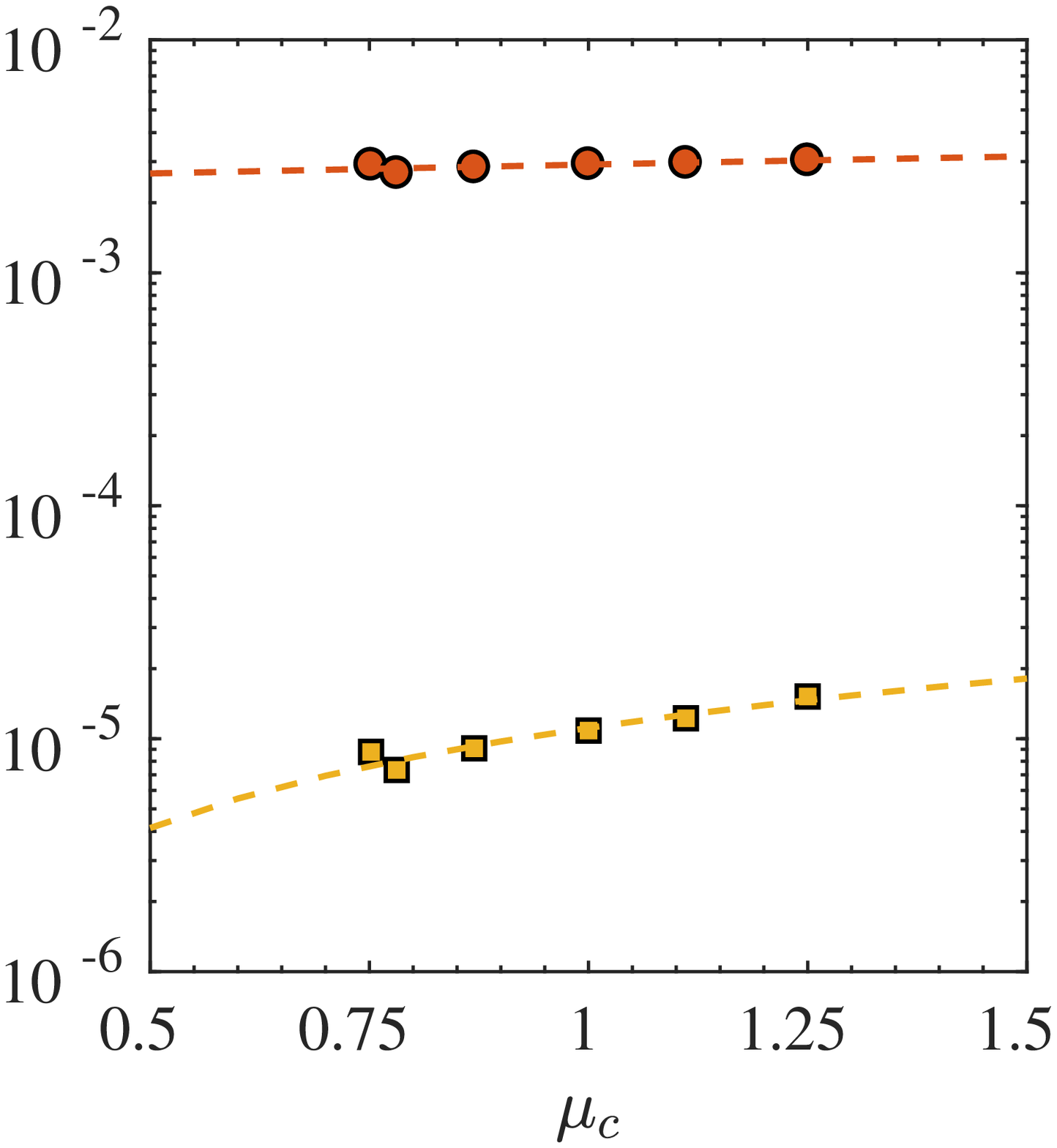}}
\caption{Average values of $E_{sw}$ and $E_{cf}$ over the integrated time history of the edge states at $Re=3000$ (a); $\overbar{Re}=3000$ (b). Dashed lines indicate linear fits of the results.}
\label{fig:Eaver}
\end{figure}

A final summary of the modified energy threshold of the edge state due to viscosity gradients is given in figure~\ref{fig:Eaver} by means of average values of $E_{sw}$ and $E_{cf}$ over the integrated time history (excluding the initial transient up to $t=4000$) for $Re=3000$ and $\overbar{Re}=3000$, and several values of the centerline viscosity.
Regardless of the choice of reference viscosity, the average energy decreases if $\mu_c<1$ and increases if $\mu_c>1$.

\begin{figure}
\centering
\includegraphics[height=0.4\textwidth]{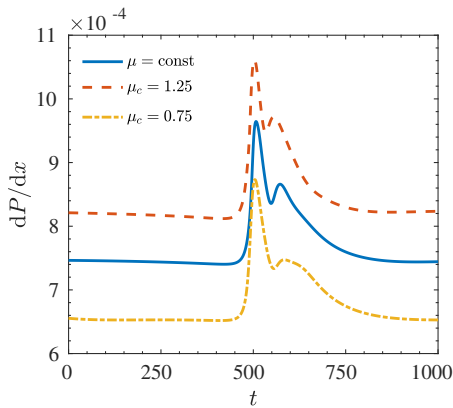}
\caption{Portion of the time history of the pressure gradient used to drive the flow at $\overbar{Re}=3000$ for $\mu_c=0.75, \, 1, \, 1.25$, while keeping the bulk velocity constant.}
\label{fig:dPdx}
\end{figure}

An alternative measure of the threshold needed to maintain the edge state on the laminar/turbulent boundary is given by the streamwise pressure gradient, which on average is directly related to the wall shear stress in non-dimensional form as
\begin{equation}\label{eq:tauwp}
\tau_w = -\overbar{Re}\frac{\mathrm{d}P}{\mathrm{d}x}.
\end{equation}
As discussed in \S\ref{sec:numerics}, $\mathrm{d}P/\mathrm{d}x$ is adapted at each time step in order to keep a constant Reynolds number based on the bulk velocity.
Figure~\ref{fig:dPdx} reports the pressure gradient calculated based on the average (in time and between upper and lower wall) wall shear stress for flow cases with constant and variable viscosity at the same $\overbar{Re}=3000$.
The relative increase of the pressure driving force averaged in time is quantified in $\Delta (\mathrm{d}P/\mathrm{d}x) = 9.8\%$ if $\mu_c=1.25$ and $\Delta (\mathrm{d}P/\mathrm{d}x) = -10.5\%$ if $\mu_c=0.75$.
The energy input required to sustain the edge state is thus smaller if viscosity decreases away from the walls, in support to its smaller energy level and reduced threshold for transition in its vicinity in the state space.

\subsection{Perturbation kinetic energy budget}
\label{sec:budget}

We define an evolution equation for the perturbation kinetic energy of a parallel flow with variable viscosity as
\begin{multline}
\label{eq:PKE}
\frac{D\langle k \rangle}{Dt} = - \frac{\partial \langle p' u'_i \rangle }{\partial x_i} + \left\langle p' \frac{\partial u'_i}{\partial x_i} \right\rangle - \frac{1}{2}\frac{\partial \langle u'_iu'_iu'_j\rangle}{\partial x_j} + \frac{1}{2}\frac{\partial}{\partial x_j} \left[ \mu_d(y) \frac{\partial \langle u'_iu'_j \rangle}{\partial x_j} \right] \\
\underbrace{-\langle u'_i u'_j \rangle \frac{\partial \langle U_i \rangle}{\partial x_j}}_{P_k} \underbrace{- \mu_d(y) \left\langle \frac{\partial u'_i}{\partial x_j}\frac{\partial u'_i}{\partial x_j} \right\rangle}_{\varepsilon_k},
\end{multline}
with the prime indicating the fluctuations with respect to the mean values calculated by using the Reynolds average, indicated by the angle brackets.
The perturbation kinetic energy is expressed as $\langle k \rangle = \frac{1}{2} \langle u'_i u'_i \rangle$, $P_k$ denotes the production and $\varepsilon_k$ the dissipation, see for example~\citet{2012_ZontaEtAl_JFMa}.
In a fully developed and statistically converged turbulent flow the balance of the right-hand-side terms of equation~(\ref{eq:PKE}) integrated across the channel height is null.
Similarly, the budget of a periodic orbit over $t = [t_0,t_0+T]$ is identically zero in order to satisfy $E_{tot}(t_0)=E_{tot}(t_0+T)$.
The terms written in divergence form are responsible for redistribution of energy, while the leading contributions to the balance are given by $P_k$ and $\varepsilon_k$.
We studied the production and dissipation balance for several periodic edge states at three different Reynolds numbers, namely $Re=2608, 3000, 3618$, and we found the same qualitative trends. 
For the sake of conciseness, we limit our discussion to $Re=2608$, as at this Reynolds number all variable and constant viscosity solutions are periodic.

\begin{figure}
\centering
\includegraphics[width=0.475\textwidth]{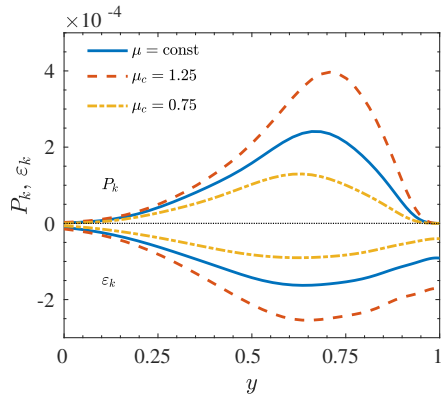}
\caption{Perturbation kinetic energy production and dissipation over one period of the edge state at $Re=2608$. The wall is located at $y=1$.}
\label{fig:pke_bud}
\end{figure}

Figure~\ref{fig:pke_bud} shows the perturbation kinetic energy production and dissipation profiles as functions of the wall normal location in the upper half of the channel, where the edge state is localised.
Due to the relatively large size of the flow structures recurring in the edge state evolution and their distance from the wall, we use outer scaling to visualise the results in a consistent fashion with the rest of the paper.
If the viscosity of the fluid increases towards the centerline, the PKE production peak increases and moves closer to the wall.
Dissipation also increases in magnitude in order to balance the larger production and to keep the edge state at the same kinetic energy level at the end of the period.
The opposite modulation of $P_k$ and $\varepsilon_k$ occurs in case viscosity decreases towards the centerline.
The described behaviour is consistent with the higher total energy in the edge state if $\mu_c>1$; the more energetic structures needed to stay on the laminar/turbulent boundary require more PKE production to be sustained over the periodic recurrence the edge state.
The opposite holds if $\mu_c<1$.

\begin{figure}
\centering
\subfigure[]{\includegraphics[height=0.375\textwidth]{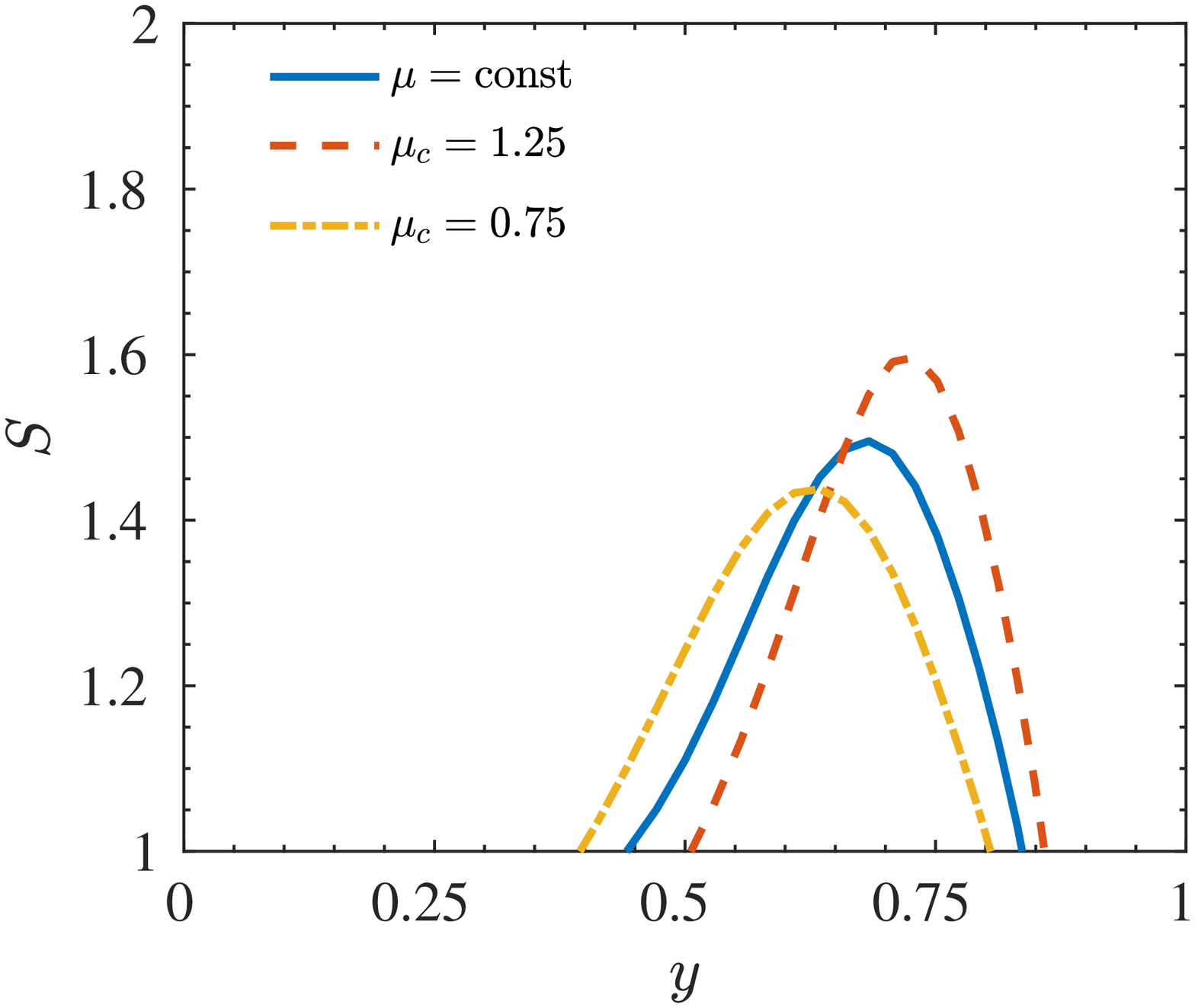}\label{fig:Ses}}
\subfigure[]{\includegraphics[height=0.373\textwidth]{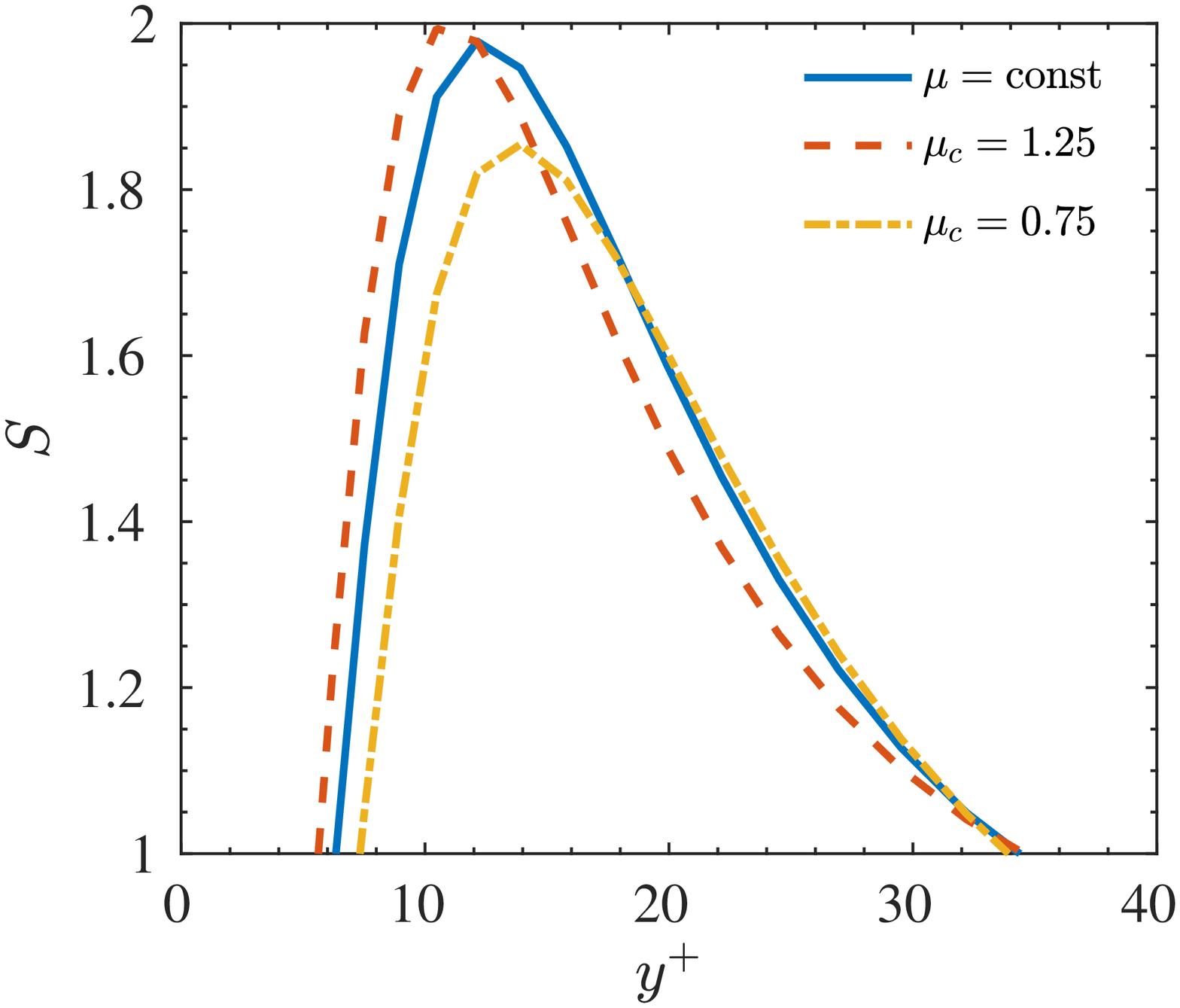}\label{fig:Stu}}
\caption{Local shear rate $S = P_k/\varepsilon_k$ (a) over one period of the edge state at $Re=2608$ and (b) in a turbulent channel flow at $Re=5000$. In (b), inner units are calculated using the friction Reynolds number of the constant viscosity case $Re_\tau = 208$. The wall is located at (a) $y=1$ and (b) $y^+=0$.}
\label{fig:S}
\end{figure}

The production and dissipation terms can be combined to have an indication on the degree of anisotropy of the fluctuating motions and on the formation of streaks by defining the so-called local shear rate $S = P_k / \varepsilon_k$, which measures the local relative importance of production over dissipation.
A critical condition for the appearance of streaks is $S>1$~\citep{1992_LamAndBanerjee}; the larger the value of $S$, the more persistent the streaks.
Figure~\ref{fig:Ses} shows the profiles of local shear rate in the upper half channel for $\mu_c = 0.75, 1, 1.25$.
In case $\mu_c>1$, the increase of $S$ compared to the case with constant viscosity is consistent with the formation of stronger streaks highlighted in \S\ref{sec:energy}.
The extent of the region of the channel in wall-normal direction where streaks form moves closer to the wall and shrinks with respect to the constant viscosity case.
The opposite effect is observed if $\mu_c<1$.

In order to assess whether the highlighted trend of local shear rate is not limited to the transitional regime but, being a relative measure of production and dissipation hence independent from their actual values, can be representative of a weakly turbulent flow, we present in figure~\ref{fig:Stu} data relative to DNS simulations of turbulent channels at $Re=5000$.
The same frozen viscosity distribution and Reynolds number definition used in discussing edge states is adopted for the turbulent cases.
Simulations are performed on a $\pi \times 2 \times \pi/2$ domain with a resolution of $64\times129\times64$ Fourier--Chebyshev--Fourier modes.
As for the simulations used to track the edge states, the flow is driven by a pressure gradient that changes in time in order to keep the bulk Reynolds number constant.
This corresponds to a friction Reynolds number $Re_\tau=208$ for the constant viscosity case.
Results for the turbulent channel show the same modulating effect discussed for the edge states; the peak of $S$ increases and moves closer to the wall if $\mu_c>1$ and the opposite occurs for $\mu_c<1$.
We can assess to what extent the initial assumption of frozen viscosity distribution discussed in \S\ref{sec:numerics} holds as the fluctuating field increases in intensity and reaches a turbulent state.
A comparison between figure~\ref{fig:Stu} and the results presented by~\citet{2012_ZontaEtAl_JFMa} for a simultaneously heated and cooled channel flow of water at $Re_\tau=180$ reveals the same qualitative modulation of $S$ in the near wall region.
Indeed, they find larger peak that moves closer to the wall where it is heated (lower viscosity at the wall, corresponding to $\mu_c>1$ in figure~\ref{fig:Stu}) and the opposite effect on the cold wall.

\subsection{Temporal recurrence and stability of edge states}
\label{sec:period}

\begin{figure}
\centering
\subfigure[]{\includegraphics[height=0.35\textwidth]{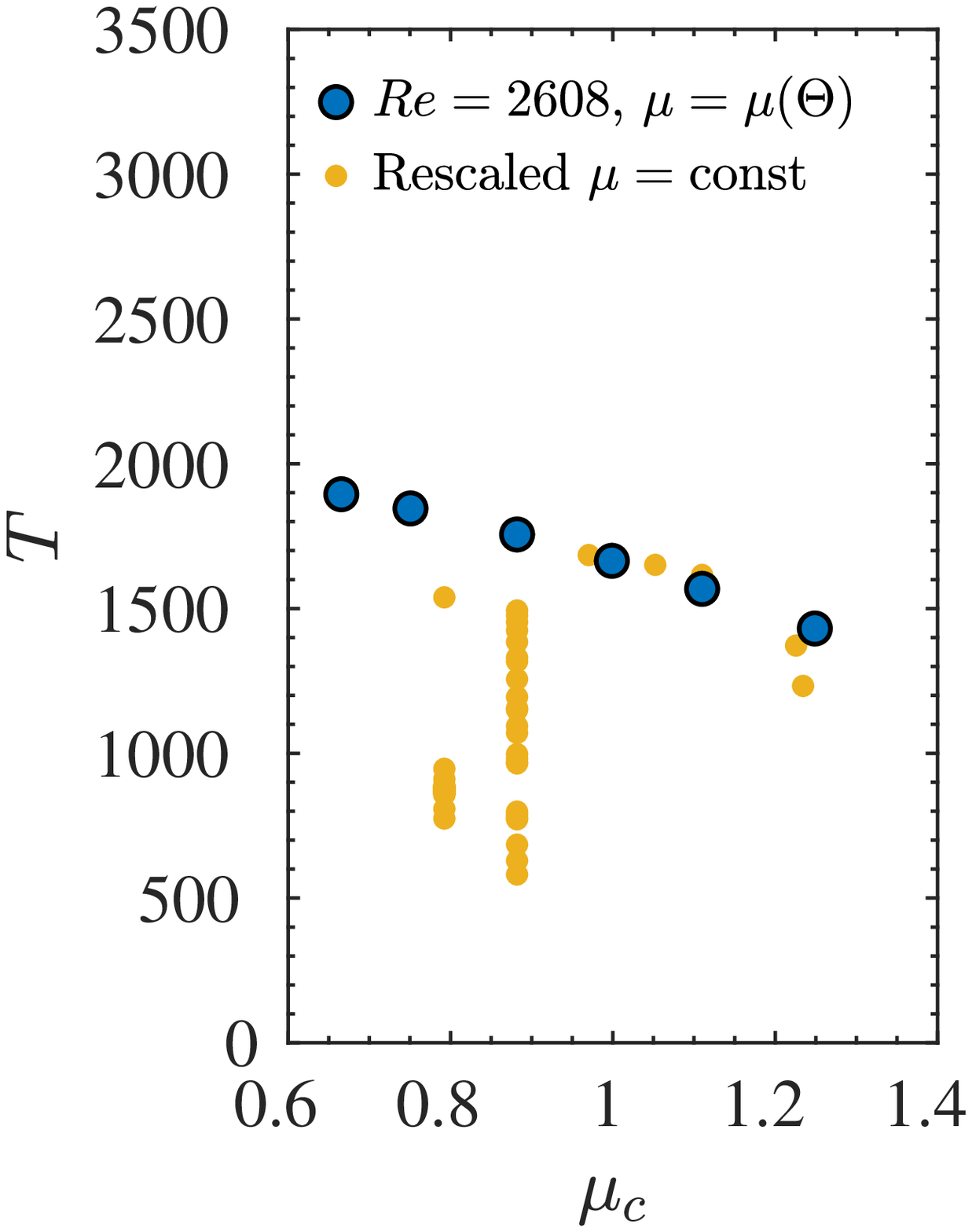}\label{fig:muc-T_2608}}
\subfigure[]{\includegraphics[height=0.35\textwidth]{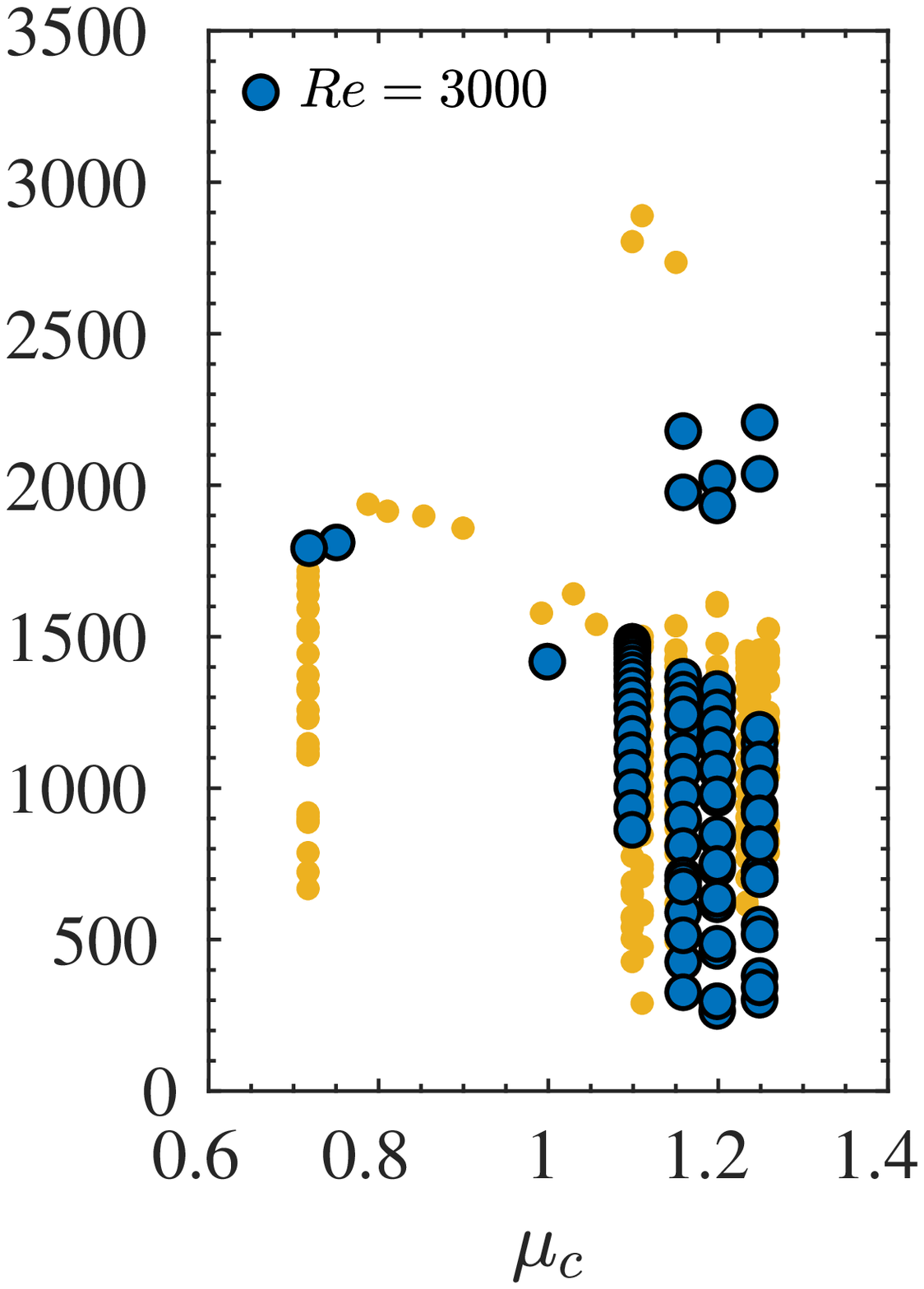}\label{fig:muc-T_3000}}
\subfigure[]{\includegraphics[height=0.35\textwidth]{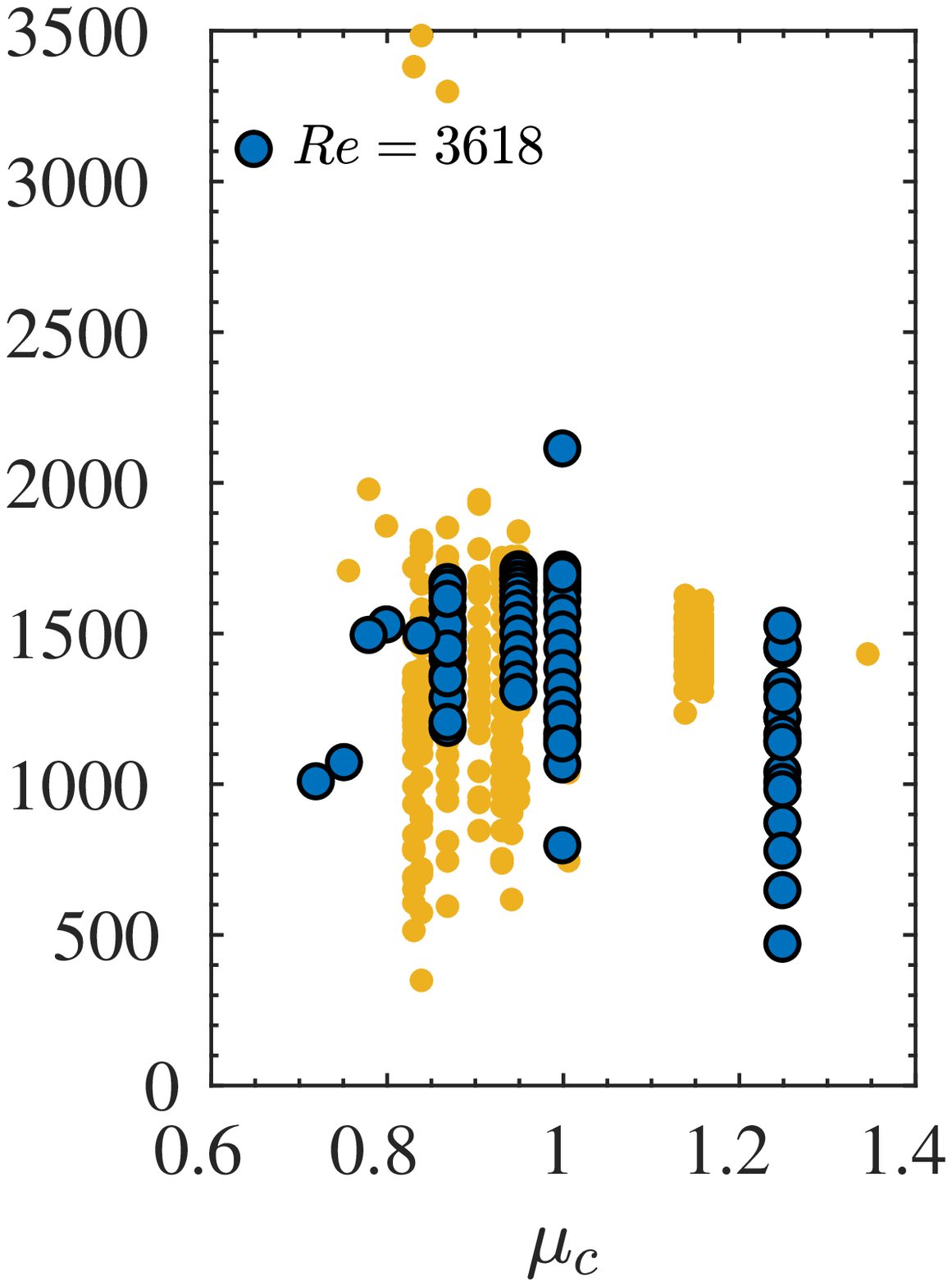}\label{fig:muc-T_3618}}
\caption{Bifurcation diagrams of the inter-burst period $T$ function of $\mu_c$ at (a) $Re=2608$; (b) $Re=3000$; (c) $Re=3618$. Dark symbols indicate variable viscosity edge states, including the reference case $\mu_c=1$.}
\label{fig:muc-T}
\end{figure}

The effect of viscosity gradients on the recurrence of the self-sustained cycle is calculated as the time interval between two consecutive burst events and is quantified in figure~\ref{fig:muc-T}, where the bifurcation parameter is the centerline viscosity, for three different Reynolds numbers, namely $Re=2608, 3000, 3618$.
Viscosity acts as a time-modulator of the edge state dynamics and, in particular, of the development of the streamwise instability of the streaks.
We highlight three effects caused by viscosity, namely i) the orbit remains periodic with a modified frequency of streak break-up, see figure~\ref{fig:muc-T_2608}; ii) a periodic state at constant viscosity can be disrupted and driven to a chaotic one by bifurcations induced by viscosity gradients, see figure~\ref{fig:muc-T_3000}; iii) a chaotic state at constant viscosity can be stabilised into a periodic orbit, see figure~\ref{fig:muc-T_3618}.
As for \S~\ref{sec:energy}, the figures additionally include rescaled results of constant viscosity flow cases.
The period $T$ is also rescaled by the factor $Re/\overline{Re}$ in order to have the same normalisation in terms of reference velocity.

In the first scenario, see figure~\ref{fig:muc-T_2608}, increasing viscosity away from the walls ($\mu_c>1$) shortens the inter-burst period, while decreasing viscosity away from the walls ($\mu_c<1$) results in the opposite effect. 
The bursting frequency only is modulated while the stability of the $\mu_c=1$ edge state is preserved, namely all the orbits are periodic.
Comparison to constant viscosity simulations shows that, even though some overall trend can be captured by rescaling $T$, see also figures~\ref{fig:muc-T_3000} and~\ref{fig:muc-T_3618}, there are significant quantitative as well as qualitative differences attributed to the effect of a wall-normal viscosity gradient, e.g. the stability of the edge state at $\mu_c=0.8$ and $\mu_c=0.88$.
The physical interpretation of increased $T$ is that the streaks instability takes more time to develop and that the two dimensional base profile is more stable.
In support of this statement, we performed a secondary stability analysis of the essentially two dimensional ($z-y$) velocity profile extracted when the streaks attain their maximum amplitude, at $t_B$ (see figure~\ref{fig:timeplot}).
Equations were discretised in the wall normal direction with 97 Chebyshev modes. The fundamental secondary instability mode was expanded in 8 spanwise Fourier modes. The streamwise wavenumber was set to $\alpha=2$ in order to match the wavelength in the DNS box. Velocities were made non-dimensional using the reference velocity at constant viscosity.
The imaginary part of the unstable mode for constant and variable viscosity cases at $Re=2608$ is reported in table~\ref{tab:sigmai} and shows that the linear growth of the instability is faster if $\mu_c>1$.

\begin{figure}
\centering
\subfigure[$Re=3000$]{\includegraphics[width=0.425\textwidth]{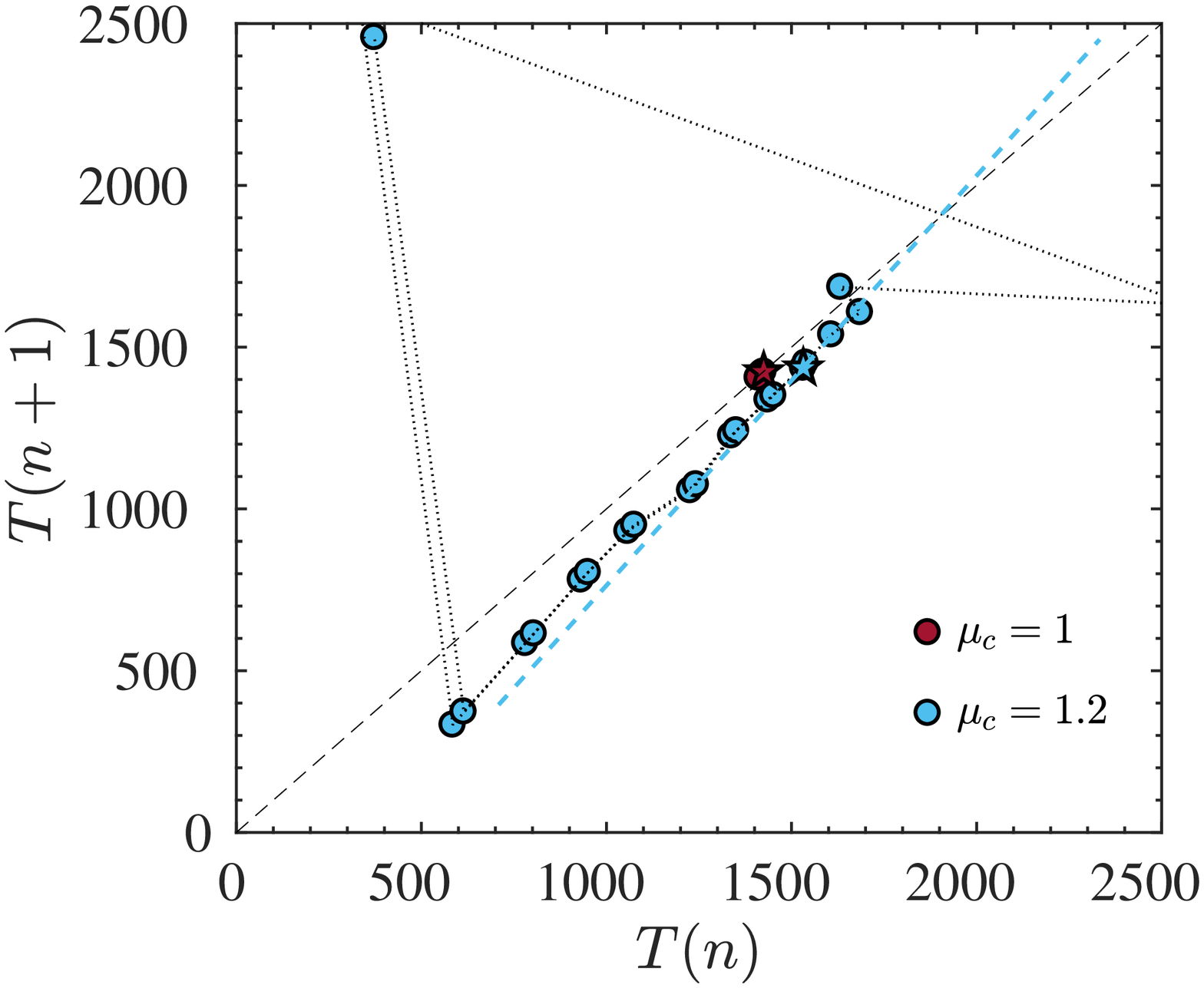}\label{fig:T-RetMap_3000}}
\subfigure[$Re=3618$]{\includegraphics[width=0.425\textwidth]{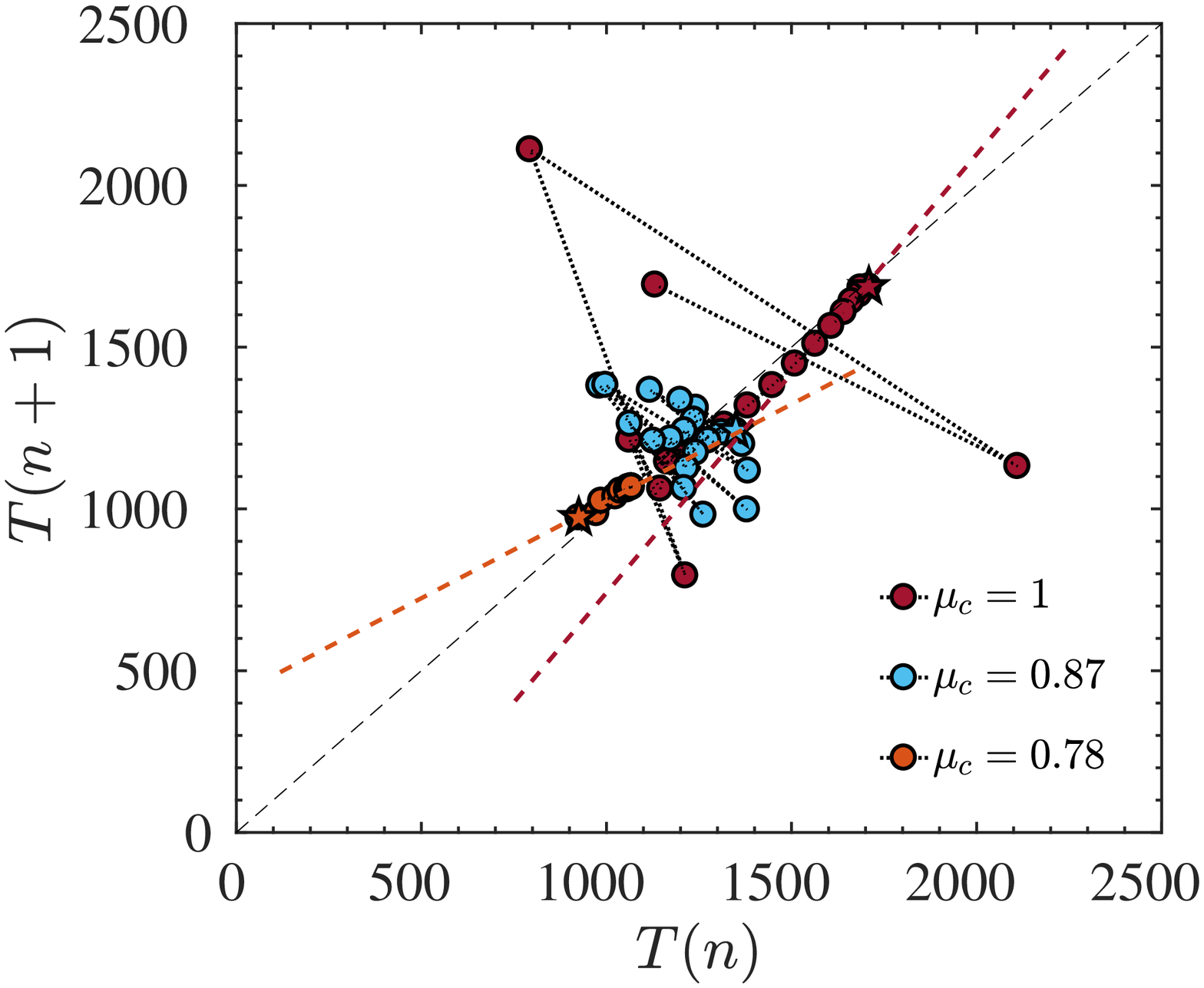}\label{fig:T-RetMap_3618}}
\caption{First return maps of the inter-burst period for selected variable viscosity cases at (a) $Re=3000$; and (b) $Re=3618$. The star symbol indicates the first pair of periods calculated after discarding the initial transient; subsequent pairs are indicated by the circles and connected by a dotted line. The dashed lines indicate the threshold for stability $T(n)=T(n+1)$ and the slopes with which the edge states approach it.}
\label{fig:T-RetMap}
\end{figure}

Particularly interesting from a dynamical system standpoint are the latter two scenarios as viscosity does not only modulate the frequency of the edge state but additionally acts on its stability.
That is conveniently assessed by means of the first return map in figure~\ref{fig:T-RetMap_3000}, in which every entry has on the horizontal axis the period of the regeneration cycle $n$ and on the vertical axis the period of the subsequent cycle $n+1$.
The first pair of periods, $n=1$ and $n=2$, are indicated by a star; the following pairs are circles connected by a dotted line.
Edge states that approach the $T(n)=T(n+1)$ line with a slope smaller than one are stable and eventually converge to a relative periodic orbit with a constant period.
On the other hand, edge states characterised by a slope larger than one are linearly unstable and aperiodic~\citep{2014_KhapkoEtAl}.
At $Re=3000$, the edge state for the constant viscosity flow is periodic with $T\simeq1400$.
Increasing viscosity towards the centerline results in the loss of stability of the edge state that becomes chaotic, see figure~\ref{fig:muc-T_2608} for the $\mu=1$ and $\mu_c=1.2$ cases.
At $Re=3618$, edge tracking for a flow with constant viscosity results in a chaotically bursting orbit, with periods fluctuating between $T=[750,2100]$.
Decreasing the centerline viscosity reduces the range of attained $T$ and, for small enough $\mu_c$, results in a linearly stable edge state orbit with constant period for $\mu_c<0.84$, see~\ref{fig:muc-T_3618}.
Figure~\ref{fig:T-RetMap_3618} reports the cases $\mu=1$; $\mu_c=0.87$, for which the edge state is chaotic as in case of constant viscosity; $\mu_c=0.78$ for which the edge state has a constant period.

\begin{table}
\centering
\caption{Imaginary part of the unstable mode predicted by a linear secondary stability calculation over the two-dimensional velocity profile at the maximum streaks amplitude $t_B$.}
\begin{tabular}{l c c c}
& $\mu_c=0.75$ & $\mu_c=1$ & $\mu_c=1.25$\\
$\sigma_i$ & $7.77\times 10^{-3}$ & $3.84\times 10^{-2}$ & $4.74\times 10^{-2}$
\end{tabular}
\label{tab:sigmai}
\end{table}

\clearpage
\section{Conclusions}
\label{sec:discussion}

We have assessed how mean viscosity gradients affect edge state solutions and how the local threshold for transition to turbulence changes in their neighbourhood of the state space.
We performed direct numerical simulations of nearly minimal flow units with a frozen symmetric viscosity profile and compared the perturbation kinetic energy of the edge states to the ones calculated for canonical flows with constant viscosity in the same domain.
In order to exclude that our discussion is affected by the arbitrary choice of the reference scale for viscosity, we compared flow cases using two different definitions of the Reynolds number, respectively based on the viscosity at the wall and on its average value across the channel.

Consistently over a range of subcritical Reynolds numbers, decreasing viscosity away from the walls results in an edge state dynamics that is sustained on an overall lower perturbation kinetic energy level with a smaller driving force quantified in terms of streamwise pressure gradient, if compared to a flow with constant viscosity. 
The perturbation kinetic energy budget over one period of the edge state regeneration cycle shows that the production term becomes smaller and that the local production-to-dissipation ratio reduces compared to a constant viscosity flow, in support of the weaker velocity streaks found and in analogy with fully turbulent flows at low Reynolds number.
The documented results indicate a decreased nonlinear stability limit for perturbations evolving in the proximity of the edge state.
Opposite conclusions in terms of perturbation kinetic energy modulation are drawn if viscosity increases away from the walls.
When comparing flows using a Reynolds number based on the average viscosity, we have shown that the applicability of a rescaling of the results of constant viscosity simulations qualitatively predicts the effect of variable viscosity on the streamwise flow structures while fails in terms of cross-flow motion, thereby highlighting the importance of fully accounting for the spatial non-uniformity of viscosity.

The results discussed in this paper suggest that the effect of spatially varying viscosity is a shift of the position of the edge state in the state space relative to the laminar attractor and turbulent saddle.
This is sketched in figure~\ref{fig:sketch0}, where the edge state moves closer to the laminar attractor in case viscosity decreases away from the walls, and vice versa.
As a consequence, the stable manifold in the vicinity of the edge state is modulated accordingly (solid line).
This state space modulation implies a reduction of the basin of attraction of the laminar state in the proximity of the edge state if viscosity is larger at the walls, and a reduced threshold for transition for perturbations that evolve in the neighbourhood of the edge state.
The interpretation of the state space role of viscosity is the opposite for flows with lower viscosity at the walls. 
Due to the limitation of the edge tracking algorithm in describing the manifold away from its relative attractors (dashed line in figure~\ref{fig:sketch0}), we can only speculate that the effect of viscosity discussed here applies also far from the edge state.
In doing so, we base our intuition on previous results from linear stability theory, which predicts a stabilisation or destabilisation of the laminar flow consistent with the analysis presented~\citep{1996_WallAndWilson,2007_SameenAndGovindarajan}, and on evidence of the increase of the Reynolds number at which nonlinear travelling waves appear in pipe flow of shear-thinning fluids~\citep{2010_RolandEtAl}, although we stress the fact that the interaction between velocity and viscosity fluctuations in the latter case is substantially different from the one in a fluid where viscosity depends on temperature only.

\smallskip
E.R. gratefully acknowledges Yohann Duguet for discussions.
Financial support from the Linn\'e FLOW  Centre is gratefully acknowledged. The simulations were performed on resources provided by the Swedish National Infrastructure for Computing (SNIC)  at HPC2N (Abisko). 
\bibliographystyle{jfm}
\bibliography{references}

\end{document}